\pdfoutput=1
\documentclass[12pt,a4paper]{article}
\usepackage{jheppub}

\usepackage{axodraw4j}
\usepackage{dcolumn}  
\usepackage{bm}    
\usepackage{amssymb} 
\usepackage{amsmath,bm}
\usepackage{amsfonts}    
\usepackage{slashed}  
 \usepackage{relsize}
\usepackage[mathscr]{euscript}
\usepackage{bbm}
\usepackage[utf8]{inputenc}

\usepackage{euscript}
\usepackage{mathtools}
\usepackage[sf,isu]{caption}
\usepackage{appendix}
\usepackage{float}
\usepackage{simpler-wick}
\usepackage{setspace}
\usepackage[export]{adjustbox}
 \usepackage{subcaption}
 \usepackage{float}
 \usepackage[english]{babel}
 \usepackage{xcolor}

\def\({\left(} \def\){\right)}
\def\[{\left[} \def\]{\right]}
\def\al{\alpha} 
\def\del{{\partial}}

\newcommand{\non}{\nonumber \\}

\newcommand{\be}{\begin{equation}}
\newcommand{\ee}{\end{equation}}
\newcommand{\bea}{\begin{eqnarray}}
\newcommand{\eea}{\end{eqnarray}}
\newcommand{\ba}{\begin{eqnarray}}
\newcommand{\ea}{\end{eqnarray}}
\newcommand{\nn}{\nonumber \\}

\newcommand{\beq}{\begin{equation}}
\newcommand{\eeq}{\end{equation}}
\newcommand{\beqa}{\begin{eqnarray}}
\newcommand{\eeqa}{\end{eqnarray}}
\newcommand{\beqar}{\begin{eqnarray*}}
\newcommand{\eeqar}{\end{eqnarray*}}
\newcommand{\e}{\epsilon}

\newcommand{\eg}{{\it e.g.,}\ }
\newcommand{\ie}{{\it i.e.,}\ }

\definecolor{labelkey}{rgb}{0.0,0.2706,0.4941}

\title{The bi-conical vector model at $1/N$}

\author[]{Noam Chai}
\author[]{\!, Eliezer Rabinovici}
\author[]{\!, Ritam Sinha}
\author[]{\! and Michael Smolkin}

\affiliation[]{\vspace{.3cm} The Racah Institute of Physics, The Hebrew University of Jerusalem \\
	Jerusalem, 91904,\,Israel \\ \vspace{-.3cm}
	}

\emailAdd{noam.chai@mail.huji.ac.il}
 \emailAdd{eliezer@mail.huji.ac.il}
 \emailAdd{ritam.sinha@mail.huji.ac.il}
  \emailAdd{michael.smolkin@mail.huji.ac.il}
\abstract{
We study finite $N$ aspects of the $O(m)\times O(N-m)$ vector model with quartic interactions in general $2\leq d \leq 6$ spacetime dimensions. This model has recently been shown \cite{Chai:2020zgq,Chai:2020onq} to display the phenomenon of persistent symmetry breaking at a perturbative Wilson-Fisher-like fixed point in $d=4-\epsilon$ dimensions. The large rank limit of the bi-conical model displays a conformal manifold and a moduli space of vacua. We find a set of three double trace scalar operators that are respectively irrelevant, relevant and marginal deformations of the conformal manifold in general $d$. We calculate the anomalous dimensions of the single and multi-trace scalar operators to the first sub-leading order in the large rank expansion. The anomalous dimension of the marginal operator does not vanish in general, indicating that the conformal manifold is lifted at finite $N$. In the case of equal ranks we are able to derive explicitly the scaling dimensions of various operators as functions of only $d$.

}

\def\f{\frac}
\def\p{\partial}
\def\nn{\nonumber}
\def\lan{\langle}
\def\ran{\rangle}

\def\l{\lambda}

\def\s{\sigma}

\def\eps{\epsilon}

\def\g{\gamma}

\def\ba{\bar{a}}

\def\bh{\textbf{h}}

\def\det{\text{det}}
\def\tr{\text{tr}}
\def\e{\mbox{\footnotesize{eff}}}

\begin{document}

\maketitle

\makeatletter
 \g@addto@macro\bfseries{\boldmath}
 \makeatother

\section{Introduction}

Conformal field theories (CFTs) have been extensively studied over the decades due to their applications in condensed matter physics, statistical mechanics and string theory. The conformal bootstrap \cite{Parisi:1972zm,Polyakov:1974gs,Ferrara:1973yt,ElShowk:2012ht,Simmons-Duffin:2016gjk}, see also \cite{Mack:2019akh} and references therein, is one of the approaches to unravel the structure of a CFT. Rather than using a specific Lagrangian, it works with general building blocks, such as the scaling dimensions of local operators and their operator product expansion coefficients. In this paper we adopt the diagrammatic $1/N$ conformal perturbation theory \cite{Vasiliev:1981yc,Vasiliev:1981dg} to explore a CFT data of the new class of critical $O(m)\times O(N-m)$ vector models recently constructed in \cite{Chai:2020zgq,Chai:2020onq}. 

\noindent
The large-$N$ (rank) limit of these models, which go by the name of bi-conical models, displays a conformal manifold, a moduli space of vacua, and a deformed moduli spaces of vacua at finite temperature. Moreover, it turns out that some of the internal symmetries of the bi-conical model are broken at arbitrary finite temperature. That said, studies in the vicinity\footnote{There is also a non-supersymmetric example in $3+1$ dimensions \cite{Chaudhuri:2020xxb}.} of $3+1$ dimensions showed that while the conclusion regarding persistent symmetry breaking at finite temperature withstands the $1/N$ corrections, the conformal manifold and the moduli space of vacua are lifted at finite $N$ \cite{Chai:2020zgq,Chai:2020onq}. 

\noindent
This is reminiscent of what happens with $\phi^6$ vector model in $2+1$ dimensions \cite{Bardeen:1983rv,Rabinovici:1987tf}. In the large rank limit it also exhibits a conformal manifold and a moduli space of vacua, which are lifted when 1/N corrections are considered. However, there is a substantial difference between the two models: while the interacting bi-conical model is stable below $d=4$, the non-trivial fixed point \cite{Pisarski:1982vz} of the finite rank $\phi^6$ theory has no ground state \cite{Bardeen:1983rv}. Although perturbatively in $1/N$ there are no signs of pathologies, \eg the scaling dimensions of various operators satisfy unitarity bounds \cite{Goykhman:2020ffn},\footnote{This is similar to the higher dimensional critical vector models which also exhibit a non-perturbative instability in $1/N$, see \eg \cite{Fei:2014yja,Giombi:2019upv}.}  the very presence of instability furnishes an obstacle to address the question of persistent symmetry breaking in the interacting $\phi^6$ model.

\noindent
The focus of the current paper is to explore the effect of finite rank corrections on the conformal manifold of the bi-conical model in general $2< d < 6$ dimensions. To this end, we use a unique hallmark of the conformal manifolds -  existence of exactly marginal operators in the theory. In particular, we identify such an operator in the large rank limit, and calculate its anomalous dimension assuming that $1/N$ is the smallest parameter in the problem. In addition, we evaluate the anomalous dimensions of the single and multi-trace scalar operators of the bi-conical model to order $1/N$.

\noindent
Our calculations suggest that there is a new strongly interacting CFT in $2+1$ dimensions. It does not exhibit a persistent symmetry breaking due to the Mermin-Wagner-Hohenberg-Coleman no-go theorem \cite{Mermin:1966fe,Hohenberg:1967zz,Coleman:1973ci}, but it is tractable within $1/N$ expansion and may serve as a theoretical lab to study the structure of strongly coupled CFTs in $2+1$ dimensions as well as duality in the context of Vasiliev's higher-spin gravity \cite{Klebanov:2002ja,Elitzur:2005kz,Elitzur:2007zz}.

\noindent
We find that quite generally the anomalous dimension of the marginal operator does not vanish, implying that the conformal manifold is lifted at $1/N$ order, and only a discrete number of fixed points is retained. Yet, there exists one value of $d$ ($2<d<3$) for which the anomalous dimension vanishes. The higher order corrections in $1/N$ only slightly modify this value, since the rank of the group is large by assumption. Hence, in principle, the line of fixed points might survive in this fractional number of dimensions. However, it is unclear whether there is an integer rank such that $d$ is an integer too.

\noindent
In section \ref{sec:setup} we review the definition of the bi-conical model, present its perturbative RG flow, as well as provide an evidence for the existence of conformal manifold in general dimension and infinite $N$. Then in section \ref{sec:anomdim} we calculate the anomalous dimensions of the single and multi-trace scalar operators of the model at the next-to-leading order in $1/N$. In section \ref{sec:discussion} we conduct a comprehensive discussion of our results and outline potential future directions. In Appendix \ref{appx: beta function} we calculate perturbative beta functions of the model in the context of Wilsonian RG flow. In Appendix \ref{appx:identities} we summarize various identities of the conformal perturbation theory, and use them to evaluate different Feynman graphs used throughout the text.

\section{Setup}
\label{sec:setup}
In this paper, we study the critical $O(m)\times O(N-m)$ vector model in $2\leq d \leq 6$ dimensions, described by the following Euclidean action\footnote{In \cite{Rychkov:2018vya,Osborn:2020cnf,Hogervorst:2020gtc} a general class of closely related critical models was analyzed at lowest order in the $\epsilon$ expansion. Earlier works on this topic are reviewed in \cite{Pelissetto:2000ek,AAharony:2003}.}
\begin{align}
 S=\int\,d^dy\,\Big(\f12(\p \phi_i)^2 
 +\f{g_{ij}}{N} \phi_i^2 \phi_j^2\Big)
 \label{action}
\end{align}
where the flavour indices $i,j=1,2$ are summed over, $\phi_1$ and $\phi_2$ are real-valued vectors of $O(m)$ and $O(N-m)$ respectively,  and the vector indices are suppressed for brevity. For convenience let us define the following notations,
\begin{equation}
 m=N x=Nx_1,\hspace{.2cm}N-m = N(1-x)=Nx_2
\end{equation}

In the large-$N$ limit this model exhibits a line of fixed points \cite{Chai:2020onq,Chai:2020zgq}. One can illustrate it explicitly in the vicinity of $d=4$. Indeed, the perturbative beta function in $d=4-\epsilon$ dimensions is given by (see Appendix \ref{appx: beta function} for details),
 \begin{align}
     & \mu {d \al_{11}\over d\mu}=\epsilon
     \Big(- \al_{11}+ x \,\al_{11}^2+(1-x)\al_{12}^2 +\frac{8 \, \al_{11}^2}{N}\Big)  ~,\nn \\
     &\mu {d \al_{22}\over d\mu} =\epsilon
     \Big(- \al_{22} + (1-x)\al_{22}^2+x\, \al_{12}^2 + \frac{8 \, \al_{22}^2}{N}\Big) ~,
          \label{beta-fns}\\
     & \mu {d \al_{12}\over d\mu} =\epsilon\,\al_{12}
     \Big(- 1 + x\al_{11}+(1-x)\al_{22} +
     \frac{2}{N}\big(\al_{11}+\al_{22}+2\al_{12}\big)\Big) ~,\nn
\end{align}
where we introduced a set of dimensionless couplings, $\al_{ij}$, defined by $g_{ij} =2\pi^2 \epsilon \, \alpha_{ij} \mu^{\epsilon}$ with $\mu$ being a floating cut off scale. The $1/N$ terms can be dropped in the large $N$ limit. As a result, the fixed points of the flow shape a conformal manifold parametrized by $\al_{12}$,
\begin{eqnarray}
     &&\al_{11}^* = \f{1 \pm \sqrt{1-4x(1-x)\al_{12}^{*2}}}{2x}~ , \quad
     \al_{22}^*=\f{1 \mp \sqrt{1-4x(1-x)\al_{12}^{*2}}}{2(1-x)}    ~,
     \nn\\
       &&|\al_{12}^*|\le\f{1}{2\sqrt{x(1-x)}}  ~.
       \label{f.p.}
 \end{eqnarray}
In fact, the two branches are connected at the end points, so together they form a closed curve. Moreover, the matrix of critical couplings is degenerate, {\it i.e.,} $\det(\al_{ij}^*)=0$. The sub-leading $1/N$ terms lift the conformal manifold \eqref{f.p.}, and only isolated fixed points remain at finite $N$ \cite{Chai:2020onq,Chai:2020zgq}, \eg the $\mathcal{O}(N)$ invariant fixed point corresponding to $\al_{11}^*=\al_{22}^*=\al_{12}^*$ and a CFT with $\al_{12}^*=0$. The latter fixed point represents a pair of decoupled critical $\phi^4$ vector models. There is a vast literature on these CFTs, and we do not aim to further explore them in what follows. However, there is an additional fixed point with $\al_{12}^*<0$ and $\det(\al_{ij}^*)=0+\mathcal{O}(1/N)$ which has been far less studied. This CFT, save the case $x=1/2$, exhibits a symmetry breaking at all  temperatures \cite{Chai:2020onq,Chai:2020zgq}, and we explore it in the current work.

\noindent
The standard lore in RG flow provides a simple recipe for calculating the scaling dimensions of various local operators which appear in the effective action of a CFT. Starting from linearizing the RG flow equations around a given fixed point, one looks for the eigenvalues, $\omega_i$, of the matrix, $\omega_{ij}$, which represents first-order derivatives of the beta functions with respect to the couplings.  These eigenvalues are simply related to the scaling dimensions, $\Delta_i$, of the operators defined by the corresponding eigenvectors,

\begin{equation}
 \Delta_i= d +\omega_i ~.
 \label{omega}
\end{equation}

Using \eqref{beta-fns}, and dropping the $1/N$ terms, yields
 \begin{align}
     \omega_{ij} = -\epsilon~\mathbb{I}_{3\times 3} + \epsilon \left(\begin{array}{ccc}
       2x\al_{11}^* & 0 & 2(1-x)\al_{12}^*\\
     0&   2(1-x)\al^*_{22}&2x\al^*_{12}\\
     x\al^*_{12}&\quad (1-x)\al^*_{12}&  \quad x\al^*_{11}+(1-x)\al^*_{22}
     \end{array}\right) ~.
 \end{align}
The three eigenvalues of this matrix at any point on the conformal manifold \eqref{f.p.} are $(-\eps, + \eps,0)$. For $\epsilon>0$ they represent a relevant, an irrelevant and a marginal deformation of a fixed point, respectively.

\noindent
The above analysis is valid in the neighbourhood of $d=4$. It does not rely on $N$ being large or small, instead it makes use of the weak couplings only. In what follows we use the large $N$ techniques to identify various deformations of the fixed points in general $d$ and express the corresponding operators in terms of the fundamental fields.  

\noindent
Unlike the traditional perturbative approach, the large-$N$ expansion is particularly useful sufficiently far away from $d=4$, where the theory becomes strongly coupled, and the corresponding CFT cannot be studied perturbatively. In particular, it provides a method to reveal the structure of a strongly coupled CFT. In our case, we use it to explore the fate of conformal manifold in $d$ dimensions. To this end we evaluate, for instance, the anomalous dimension of the operator which happens to be exactly marginal in the large-$N$ limit. This operator shifts between the CFTs along the line of fixed points.

 \subsection{Conformal manifold in general dimension}

In the previous subsection we reviewed a perturbative analysis which reveals the existence of a conformal manifold in the vicinity of $d=4$. Here we build on the large-$N$ expansion to provide an evidence that conformal manifold exists in general $2< d < 6$, save the case $d=4$, where the model is free. 
 
\noindent 
We start from rewriting \eqref{action} in terms of auxiliary  Hubbard-Stratonovich (HS) fields, $\sigma_i$, for a comprehensive review see \cite{Moshe:2003xn}, 
 \begin{align}
   S[\s_i,\phi_i]
   &=\int\,d^dy\left(\f12(\p_{\mu}\phi_i)^2  -\f{h^{ij}}4\s_i\s_j+\f1{\sqrt{N x_i}}\s_i\phi_i^2\right)~,
   \label{eff-action}
 \end{align}
where $h^{ij}$ denotes the inverse of rescaled matrix of couplings $h_{ij}=\sqrt{x_i x_j} ~g_{ij}$. Integrating out the auxiliary fields gives back the original action of the model. 

\begin{figure}[t!]
\centering
    \begin{subfigure}[t]{0.4\textwidth}
    	\centering
   	 \includegraphics[scale=0.4]{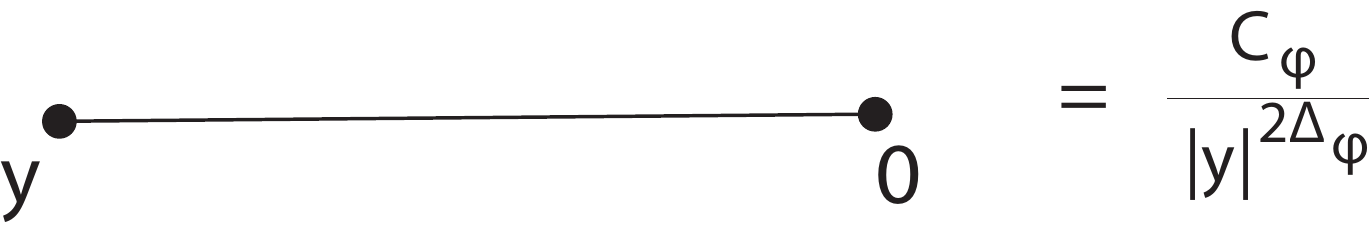}
	 \caption{}
    \end{subfigure}
    \hfill
     
     \hfill
      \begin{subfigure}[t]{0.4\textwidth}
      	\centering
    	\includegraphics[scale=0.4]{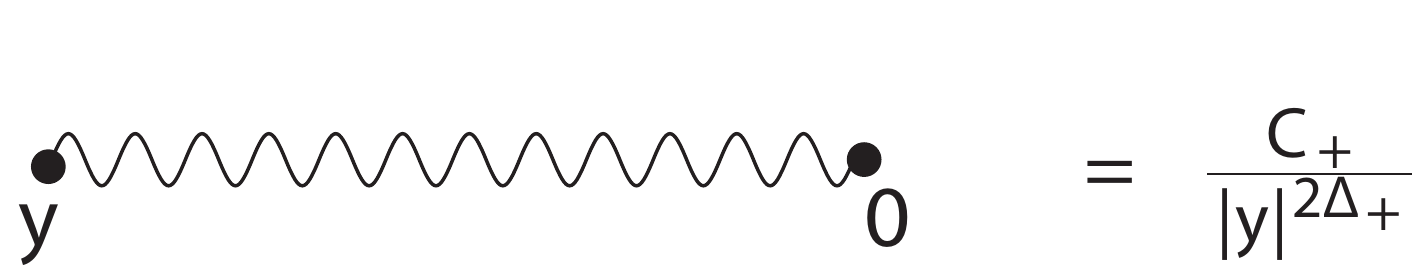}
	\caption{}
    \end{subfigure}
    \hfill
     \begin{subfigure}[t]{0.4\textwidth}
     	\centering
    	\includegraphics[scale=0.4]{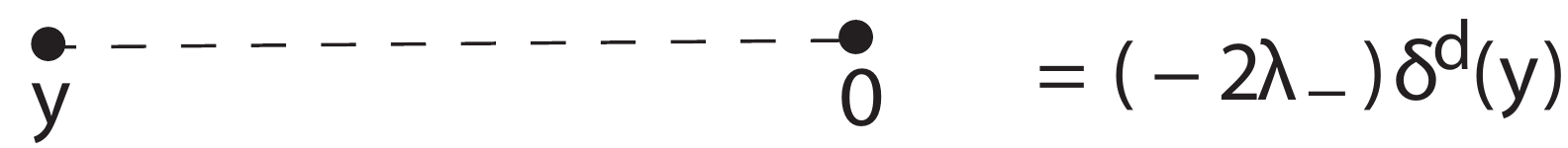}
	\caption{}
    \end{subfigure}
  
    \caption{Feynman rules for various propagators in the model. }
    \label{feyn-rules-1}
 \end{figure}

\noindent
The large-$N$ propagators of $\phi_i$'s are
\be
 \lan\phi_i(x)\phi_j(y)\ran =\f{C_\phi}{|x-y|^{2\Delta_\phi}}\delta_{ij}~, \quad \Delta_\phi=\f{d}2-1~,
 \quad C_\phi = {\Gamma\({d-2\over 2}\) \over 4\pi^{d/2}} ~,
\ee
where the vector indices are suppressed for brevity, and we used a Fourier transform relation,
\begin{equation}
\int\frac{d^dp}{(2\pi)^d}\,e^{ip\cdot y}\frac{1}{(p^2)^{\frac{d}{2}-\Delta}}=
\frac{2^{2\Delta-d}}{\pi^{\frac{d}{2}}}\frac{\Gamma(\Delta)}{\Gamma\left(\frac{d}{2}-\Delta\right)}
\frac{1}{|y|^{2\Delta}}\,.
\label{useful}
\end{equation}
The $\phi_i$ propagators will be represented by a solid line in the Feynman graphs, see Fig.\ref{feyn-rules-1} (a).

\noindent
In fact, the path integral over $\phi_i$ can be done in a closed form, because \eqref{eff-action} is gaussian in these fields, and we get
\begin{align}
 S_{\e}[\s_i] = -\f{h^{ij}}4 \int\,d^dy ~\s_i\s_j  + \f{N}2\sum_{i} x_i ~ \tr\log\Big(-\p^2 + \f2{\sqrt{N x_i}}\s_i\Big)~.
 \label{eff-action2}
\end{align}
Expanding the trace log term around the conformal vacuum, yields
\begin{align}
 S_{\e}[\s_i] &=
 -\int\,d^dy_1 d^d y_2 ~ \s_i(y_1) \Big({h^{ij}\over 4}\delta(y_1-y_2)  + \delta^{ij}  B(y_1-y_2)\Big)\s_j(y_2)+\ldots 
    \label{eff-action-2}
\end{align}
where we explicitly keep quadratic terms only, whereas ellipsis encode the rest, and\footnote{$ B (x) $ denotes the product of two free massless scalar field propagators.} 
\begin{align}
 B(y)={C_\phi^2\over |y|^{2(d-2)}}~.
 \label{bubble-int}
\end{align}
The large-$N$ effective action is gaussian, but in general it includes an infinite tower of higher order effective vertices suppressed by powers of $1/N$. For our needs, however, the infinite series of terms can be truncated at the fourth order in $\s_i$, because higher order vertices do not contribute to the leading order anomalous dimensions. 

\noindent
As usual, inverting the quadratic part of \eqref{eff-action-2} yields the matrix of propagators for the HS fields. In momentum space, it takes the form
\begin{align}
   \label{prop_matrix}
    \langle \sigma_i(p) \sigma_j(-p)\rangle &=-\f{2}{[1+4B(p) \text{Tr} (\bh)+16 B^2(p)\det(\bh)]}
    \big(h_{ij} + 4B(p)\det(\bh)\delta_{ij} \big) ~,
\end{align}
To identify the scaling dimensions of propagating fields at the fixed point, it is useful to diagonalize the matrix of couplings, $\bh$. Let us denote by $\l_\pm$ and $U^i_\pm$ its eigenvalues  and {\it normalized} eigenvectors respectively. Projecting \eqref{prop_matrix} onto $U^i_\pm$, yields the following propagators 
\bea
 \langle \sigma_+(p) \sigma_+(-p)\rangle &=& - {2\lambda_+\over 1+ 4\lambda_+ B(p)}  ~,  \quad \langle \sigma_+(p) \sigma_-(-p)\rangle=0~,
 \nonumber \\
 \langle \sigma_-(p) \sigma_ -(-p)\rangle &=& - {2\lambda_-\over 1+ 4\lambda_- B(p)} ~,
 \label{conf_prop}
\eea
where $\sigma_\pm=U^i_\pm \sigma_i$. 

\noindent
As we pointed out in the previous subsection, $\det(\bh^*)\sim\mathcal{O}(1/N)$ at any point of the conformal manifold \eqref{f.p.}. This degeneracy is true to all orders in $\epsilon$-expansion. Therefore one of the eigenvalues is suppressed relative to the other at the fixed points that we are interested in, \ie $\lambda_-\sim \mathcal{O}(1/N)$, whereas $\lambda_+\sim \mathcal{O}(1)$, because $\lambda_+\lambda_-=\det(\bh^*)\sim\mathcal{O}(1/N)$. Hence, in the deep IR (or UV) regime, defined by $N\gg B(p)\lambda_+\gg1$, in $2\leq d\leq 4$ (or $4<d\leq 6$) dimensions, the propagators simplify
\bea
 \langle \sigma_+(p) \sigma_+(-p)\rangle &=& - {1\over 2 B(p)}  ~,  \quad \langle \sigma_+(p) \sigma_-(-p)\rangle=0~,
 \nonumber \\
 \langle \sigma_-(p) \sigma_ -(-p)\rangle &=& - 2 \lambda_ - \Big( 1 - 4 \lambda_ - B(p) \Big)  ~,
 \label{conf_prop}
\eea
Note that we keep the next-to-leading order term in the propagator of $\sigma_-$, because to leading order in the large-$N$ expansion it boils down to a contact term in position space.  

\noindent
Using the Fourier transform formula \eqref{useful} one can get the large-$N$ scaling dimension of the fields based on their momentum space behaviour. Since $B(p)\sim p^{d-4}$, we find that in the large $N$ limit, $\s_+$ scales as $\Delta_+=2$. As for the $\s_-$ propagator, the first term within the parentheses in \eqref{conf_prop} corresponds to a contact term, and therefore the scaling dimension of $\s_-$ is determined solely by the last term, \ie $\Delta_- = d-2$. 

\noindent
Having $\sigma_\pm$ at hand one can construct new primary fields. Thus, for instance, in the large-$N$ limit the scaling dimensions of primary operators $\s_+^2$, $\s_-^2$ and $\s_+\s_-$ are equal to $4, 2d-4$ and $d$ respectively. Indeed, the model is gaussian in this limit, and therefore the scaling dimensions are additive under composition of operators. In particular, the above scaling dimensions match what we have found in the previous subsection using a perturbative beta function in the vicinity of $d=4$. Moreover, it can be explicitly verified that $\s_+^2$, $\s_-^2$ and $\s_+\s_-$ correspond to the eigenvectors of $\omega$. This completes our construction of the primary operators which represent admissible nearly marginal deformations of the critical model in the vicinity of $d=4$. 

\noindent
Remarkably, we showed that the large-$N$ scaling dimension of $\s_+\s_-$ equals $d$ in any number of dimensions, implying that to leading order in the $1/N$ expansion this operator is marginal. This provides an evidence for the existence of conformal manifold in general $d$. Our primary goal is to calculate the leading order anomalous dimensions of  $\s_+^2$, $\s_-^2$ and $\s_+\s_-$. If the scaling dimension of the marginal operator $\s_+\s_-$ acquires a non-trivial correction, then the line of fixed points is lifted by the finite rank corrections in any number of dimensions. 

\subsection{Feynman rules and useful identities}

We finish this section by introducing a number of Feynman rules used throughout the rest of our work. First we notice that in the large-$N$ limit the propagators of the Hubbard-Stratonovich fields in a non-diagonal basis is dominated by $\langle \sigma_+ \sigma_+\rangle$. Indeed, by definition
\be
 \langle \sigma_i(y) \sigma_j(0)\rangle=U_+^i U_+^j \langle \sigma_+(x) \sigma_+(0)\rangle
 +U_-^i U_-^j\langle \sigma_-(x) \sigma_-(0)\rangle ~,
\ee
where according to \eqref{conf_prop} the contribution of $\langle \sigma_-\sigma_-\rangle$ is $1/N$ suppressed. 

\noindent
In position space we have
\bea
  \lan\s_+(y)\s_+(0)\ran &=& \f{C_+}{|y|^{2\Delta_+}} ~, \quad 
  C_+ = \f{2^d\Gamma\left(\f{d-1}2\right)\sin\left(\f{\pi d}2\right)}{\pi^{3/2}\Gamma\left(\f{d}2-2\right)} ~,
  \nn \\
    \lan\s_-(y)\s_-(0)\ran &=& -2\lambda_- \Big(\delta(y)  -4 \lambda_- B(y) \Big) ~,
  \label{prop}
\eea
where we used \eqref{useful}. 

\begin{figure}[t!]
\centering
    \begin{subfigure}[t]{0.4\textwidth}
    	\centering
    	\includegraphics[scale=0.6]{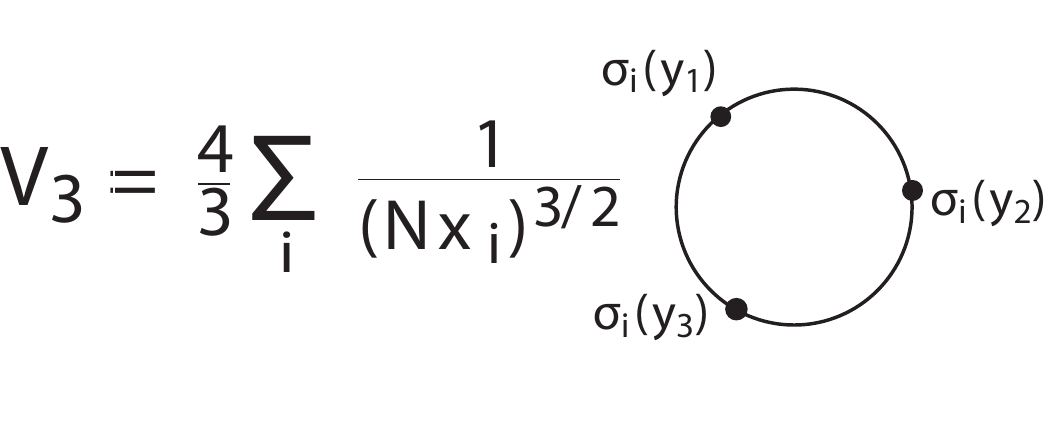}

    \end{subfigure}
    \hfill
     \begin{subfigure}[t]{0.4\textwidth}
     	\centering
      	\includegraphics[scale=0.6]{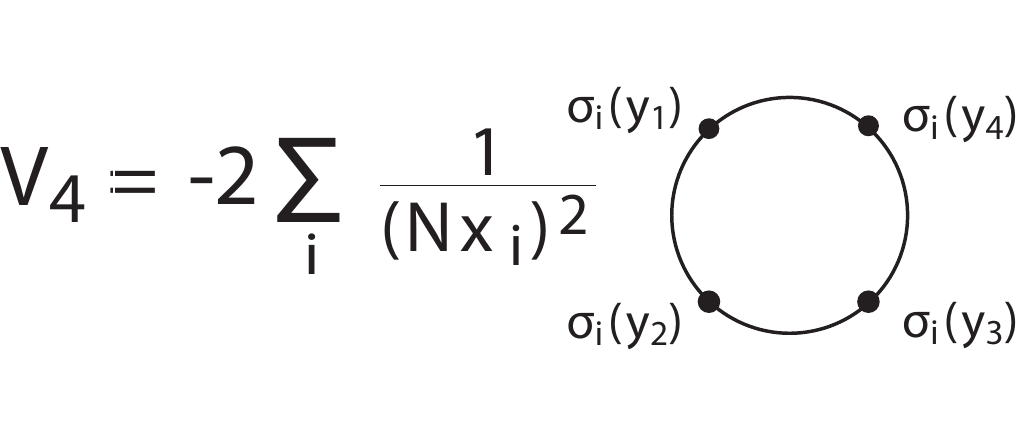}

     \end{subfigure}
    \hfill
    \caption{Effective cubic and quartic vertices obtained by expanding \eqref{eff-action2} around a conformal vacuum. Note that we implicitly sum over the vector indices of the fields $\phi_i$ which are running in the loop.}
     \label{effective_vertices}
 \end{figure}

\noindent
The $\lan\s_+ \s_+\ran$ propagator and the leading order contact term in  $\lan\s_- \s_-\ran$ will be represented by the wavy and dashed lines respectively, see Fig.\ref{feyn-rules-1} (b),(c). Expanding \eqref{eff-action2} around $\sigma_i=0$ results in an infinite series of non-local effective vertices. Two of them, namely cubic and quartic interactions, are essential for our needs. They are given by 
\be
 V_3= {4\over 3} \sum_{i=1}^2 {C_\phi^3\over \sqrt{N x_i}} \int {\prod_{j=1}^3 d^dy_j ~ \sigma_i(y_j) \over \big( |y_{12}||y_{23}||y_{31}| \big)^{d-2}} ~,
 ~
 V_4= -2 \sum_{i=1}^2 {C_\phi^4\over N x_i} \int {\prod_{j=1}^4 d^dy_j ~ \sigma_i(y_j) \over \big( |y_{12}| |y_{23}| |y_{34}| |y_{41}| \big)^{d-2}} ~,
\ee
Diagrammatically, they can be represented as in Fig. \ref{effective_vertices}.

\noindent
Furthermore, using the completeness relations, we get a useful set of identities,
\bea
 h^*_{ij}&=&\lambda_+ U_+^i U_+^j  + \lambda_- U_-^i U_-^j  \quad \Rightarrow \quad
  U_+^i U_+^j = {h^*_{ij} \over \text{Tr} (\bh^*) }+ \mathcal{O}(1/N) ~,
 \nn \\
 \delta_{ij} &=&  U_+^i U_+^j  + U_-^i U_-^j  \quad \Rightarrow \quad
 U_-^i U_-^j  = \delta_{ij} - {h^*_{ij} \over \text{Tr} (\bh^*) }  + \mathcal{O}(1/N) ~.
 \label{completeness}
\eea

\section{Anomalous dimensions}
\label{sec:anomdim}

In this section we calculate the anomalous dimensions of various operators. We use the diagrammatic $1/N$ technique, \eg \cite{Vasiliev:1975mq,Vasiliev:1981yc,Vasiliev:1981dg,Derkachov:1997ch,Gracey:2018ame,Goykhman:2019kcj}. To simplify our notation we suppress asterisk in the superscript of critical couplings. This should not result in confusion, since we never consider a non-critical model.

\noindent
Our final results passed numerous checks. For instance, in the vicinity of $d=4$, one can use the perturbative beta function \eqref{Wil_beta_func} to evaluate the scaling dimensions of the double trace operators up to $1/N$ order through the use of \eqref{omega}. We carried out such a comparison and found that the results match.

\subsection{Single trace scalars}

The full conformal two point functions of $\sigma_\pm$ have the following form
\be
 \lan \sigma_\pm(y) \sigma_\pm(0) \ran = {C_\pm(1+A_\pm) \over |y|^{2(\Delta_\pm+\gamma_\pm)}} \mu^{-2\gamma_\pm}~,
\ee
where $\mu$ represents an arbitrary floating cut off scale, $A_\pm\sim\mathcal{O}(1/N)$ are associated with the $1/N$ corrections to the leading order amplitudes $C_+$ and $C_-=8\lambda_-^2C_\phi^2$, as defined in \eqref{prop}, and $\gamma_\pm\sim\mathcal{O}(1/N)$ are the anomalous dimensions.

\begin{figure}[t!]
\centering
    \begin{subfigure}[t]{0.3\textwidth}
    	\centering
    	\includegraphics[scale=0.38]{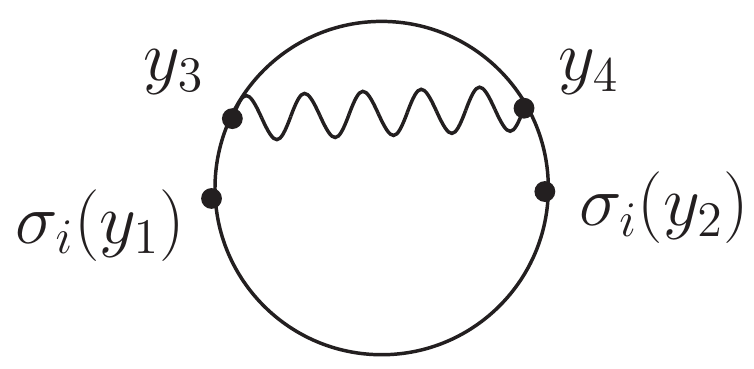}
    	\caption{}
    \end{subfigure}
    \hfill
     \begin{subfigure}[t]{0.3\textwidth}
    	\centering
      	\includegraphics[scale=0.38]{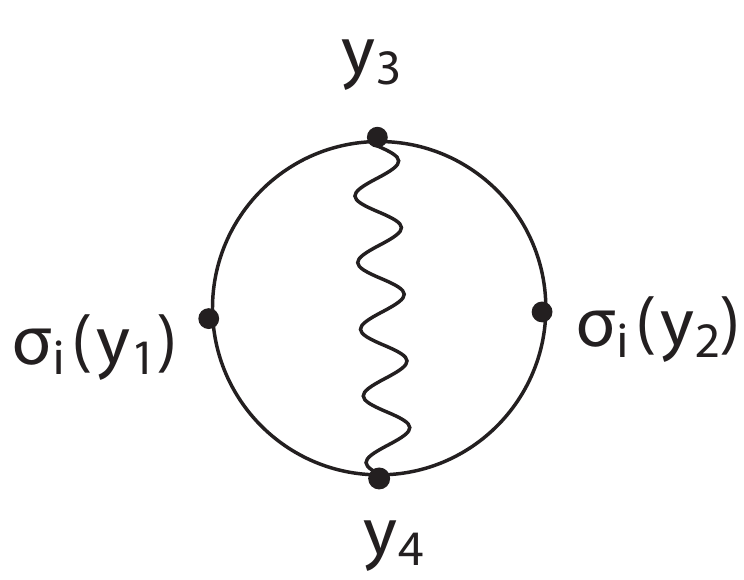}
      	\caption{}
     \end{subfigure}
    \hfill
    \begin{subfigure}[t]{0.3\textwidth}
     	\centering
      	\includegraphics[scale=0.38]{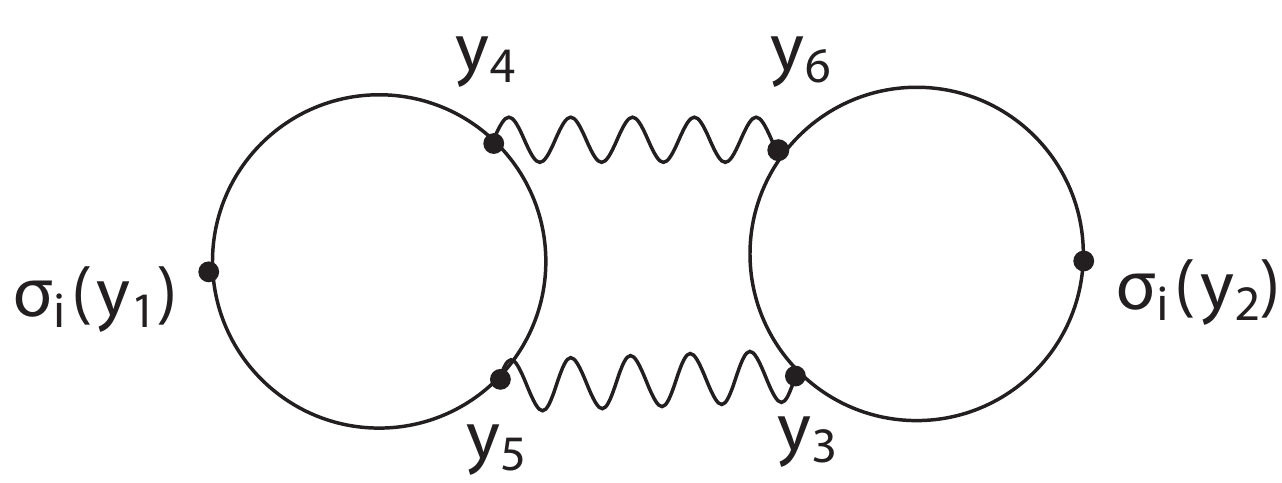}
      	\caption{}
    \end{subfigure}
    \hfill
    \caption{Feynman diagrams representing a $1/N$ correction to the effective action \eqref{eff-action-2}.}
    \label{Vsigsig}
 \end{figure}

\noindent
To derive $\gamma_\pm$ we consider $1/N$ correction, $V_2$, to the quadratic effective action \eqref{eff-action-2}. It is determined by the sum of three diagrams shown in Fig. \ref{Vsigsig},
\be
 -V_2= \text{Fig.} \ref{Vsigsig}\text{(a)} + \text{Fig.} \ref{Vsigsig}\text{(b)} 
 + \text{Fig.} \ref{Vsigsig}\text{(c)} ~.
\ee
The two-loop Feynman graphs in Fig. \ref{Vsigsig} (a),(b) are built off the quartic term in the power series expansion of \eqref{eff-action2} around the conformal vacuum, whereas the three-loop diagram in Fig. \ref{Vsigsig} (c) is obtained by contracting two cubic terms in the expansion of \eqref{eff-action2}. These effective vertices are shown in Fig. \ref{effective_vertices}. Thus, for instance,
\be
 \text{Fig.} \ref{Vsigsig}\text{(a)}=  - {4(d-2)\pi^{d/2}\over N\Gamma\({d+2\over 2}\)}  
 \sum_i {U_+^i U_+^i \over x_i } \, C_\phi^4 C_+ 
 \int d^dy_1d^dy_2 {\sigma_i(y_1)\sigma_i(y_2) \over |y_{12}|^{d-2}} \int {d^dy_3 \over |y_{13}|^{d-2}|y_{23}|^d} ~,
\ee
 where we used the following identities to integrate over $y_4$,
  \be
  {2 d \over |y|^{d+2}} = \del_\mu\del^\mu {1\over |y|^d} \quad ,  \quad\quad  \del_\mu\del^\mu {1\over |y|^{d-2}}=-{2(d-2)\pi^{d/2}\over \Gamma\({d\over 2}\)} ~\delta(y)~.
  \label{useful2}
 \ee
The integral over $y_3$ diverges in the vicinity of $y_2$. Introducing a floating spherical sharp cut off $\mu$, yields\footnote{Specifically,
 \be
  \int d^dy_3 {1\over |y_{13}|^{d-2}|y_{23}|^d} = {1\over |y_{12}|^{d-2} }\int_{1/ \mu} {d^dy_3 \over  |y_3|^d } + \ldots
  ={1\over |y_{12}|^{d-2} } {2\pi^{d/2}\over \Gamma\({d\over 2}\)} \log\mu + \ldots ~.
 \ee } 
\be
 \text{Fig.} \ref{Vsigsig}\text{(a)}=  - {16(d-2)\pi^d\over Nd\,\Gamma^2\({d\over 2}\)}  
   \sum_i {U_+^i U_+^i \over x_i } \,  C_\phi^4 C_+ 
 \int d^dy_1d^dy_2 {\sigma_i(y_1)\sigma_i(y_2) \over |y_{12}|^{2(d-2)}} \log \mu +\ldots~,
\ee
where ellipsis encode $\mu$-independent terms. Similarly,
 \bea
 \text{Fig.} \ref{Vsigsig}\text{(b)}&=&  {4\over N}  U\Big({d-2\over 2},{d-2\over 2},2\Big)
  \sum_i {U_+^i U_+^i \over x_i } \,  C_\phi^4 C_+ 
 \int d^dy_1d^dy_2 {\sigma_i(y_1)\sigma_i(y_2) \over |y_{12}|^{d-4}} \nn \\
 &&\times \int d^dy_3 {1\over |y_{13}|^d|y_{23}|^d} ~,
\eea
where we used the star-triangle identity \eqref{uniqueness scalar} to integrate over $y_4$. This time the integral over $y_3$ diverges in the vicinity of both $y_1$ and $y_2$. As before, we regulate it by introducing a spherical sharp cut off $\mu$ and retain the divergent part only, 
 \be
 \text{Fig.} \ref{Vsigsig}\text{(b)}=  {16\pi^{d/2}\over N\Gamma\({d\over 2}\)}  U\Big({d-2\over 2},{d-2\over 2},2\Big)
   \sum_i {U_+^i U_+^i \over x_i } \, C_\phi^4 C_+ 
 \int d^dy_1d^dy_2 {\sigma_i(y_1)\sigma_i(y_2) \over |y_{12}|^{2(d-2)}} \log\mu  ~.
 \ee
Finally,
  \bea
 \text{Fig.} \ref{Vsigsig}\text{(c)}&=&  {64\pi^{d/2}\over N\Gamma\({d\over 2}\)}  
 U^2\Big({d-2\over 2},{d-2\over 2},2\Big)U\Big(1,2, d-3\Big)
  \sum_{i,j}\(U_+^i U_+^j\)^2 {C_\phi^6 C_+^2 \over\sqrt{x_i x_j}} 
 \nn \\
 &&\times \int d^dy_1d^dy_2 {\sigma_i(y_1)\sigma_j(y_2) \over |y_{12}|^{2(d-2)}} \log\mu  ~,
 \eea
where the star-triangle identity \eqref{uniqueness scalar} was used to integrate over $y_6$ and $y_5$ first, and then over $y_4$. The remaining divergent integral over $y_3$ was regulated through the use of spherical sharp cut off $\mu$.
 
\noindent 
Combining all together and using \eqref{completeness} to express everything in terms of critical couplings, yields
\be
 -V_{2}=
 {4\Gamma(d-2)\sin\({\pi d\over 2}\) \over Nd(d-2)\pi^{d+1}}
  \sum_{i,j}\Big( {h_{ii}\,\delta_{ij}\over x_i  \text{Tr} (\bh)} +{d(d-3)\over 2\sqrt{x_i x_j}}  {h^2_{ij}\over \text{Tr}^2 (\bh)} \Big)
   \int \int {\sigma_i(y_1)\sigma_j(y_2) \over |y_{12}|^{2(d-2)}} \log\mu  ~.
\ee

\noindent
This correction to the effective action determines the leading order anomalous dimensions, $\gamma_\pm$, which can be read off after substituting $\sigma_i=U^i_+\sigma_+ + U^i_-\sigma_-$ and factoring out a term proportional to $B(x)$ in the leading order quadratic action \eqref{eff-action-2},
 \be
  \gamma_\pm=\pm{2^d \Gamma\({d-1\over 2}\) \sin\({\pi d\over 2}\) \over N\pi^{3/2} \Gamma\({d+2\over 2}\) }
   \sum_{i,j} U_\pm^iU_\pm^j\Big( {h_{ii}\,\delta_{ij}\over x_i  \text{Tr} (\bh)} 
  +{d(d-3)\over 2\sqrt{x_i x_j}}  {h^2_{ij}\over \text{Tr}^2 (\bh)} \Big) ~.
 \ee
Or equivalently, using the completeness relations \eqref{completeness}, yields
\bea
  \gamma_+&=&{2^d \Gamma\({d-1\over 2}\) \sin\({\pi d\over 2}\) \over N\pi^{3/2} \Gamma\({d+2\over 2}\) }
   \sum_{i,j}\Big( {1 \over x_i } {h^2_{ii}\delta_{ij}\over \text{Tr}^2 (\bh)}  
  +{d(d-3)\over 2\sqrt{x_i x_j}}  {h^2_{ij} h_{ij}\over \text{Tr}^3 (\bh)} \Big) ~, 
  \nn \\
  \gamma_- &=& -{2^d \Gamma\({d-1\over 2}\) \sin\({\pi d\over 2}\) \over N\pi^{3/2} \Gamma\({d+2\over 2}\) }
   \sum_i\Big( {h_{ii} \over x_i \text{Tr} (\bh)} 
  +{d(d-3)\over 2 }  {h^2_{ii}\over x_i \text{Tr}^2 (\bh)} \Big) + \gamma_+~.
  \label{gamma_p}
 \eea

It is hard to further simplify this expression without knowing the critical values of the coupling constants in general $d$. However, in the special case of equal ranks, $x_1=x_2=1/2$, the anomalous dimensions can be written in terms of dimension $d$ only. Indeed,  in this case the fixed points \eqref{f.p.}, which survive the $1/N$ corrections, respect the $\mathbb{Z}_2$ symmetry $\phi_1 \leftrightarrow\phi_2$, \ie they satisfy $h_{11}=h_{22}$, as can be explicitly checked within $\epsilon$-expansion. Recall that to leading order in $1/N$ these couplings are also equal to $|h_{12}|$, because $\det(\bh)\sim\mathcal{O}(1/N)$. Hence,

  \be
  \gamma_\pm\Big|_{m={N\over 2}}=\pm{2^d \Gamma\({d-1\over 2}\) \sin\({\pi d\over 2}\) \over N\pi^{3/2} \Gamma\({d+2\over 2}\) }
  \Big( 1 
  +{d(d-3)\over 4}   \big(1\pm\text{sign}(h_{12})\big) \Big)~.
  \label{sing_anom}
 \ee

 \subsection{Double trace scalars}

In this subsection, we proceed to calculate the leading order anomalous dimensions of the double trace scalars $\sigma_+^2, \sigma_-^2$ and $\sigma_+\sigma_-$.  As shown in the previous section, the scaling dimensions of these operators in the large $N$ limit are equal to $4, 2d-4$ and $d$ respectively.  Hence, the anomalous dimensions of $\sigma_+^2$ and $\sigma_-^2$ cannot change their RG behaviour in $d\neq 4$, \ie they stay relevant or irrelevant provided that $N$ is sufficiently large. However, a non-zero sub-leading correction to the scaling dimension of the marginal operator, $\sigma_+\sigma_-$, determines whether it becomes relevant or irrelevant, depending on the sign of correction. In particular, if the leading order anomalous dimension vanishes, then it implies that the line of fixed points survives up to $\mathcal{O}(1/N)$ order. In either case, the results would carry important information regarding the phase structure of the theory. 

\begin{figure}[H]
    \begin{subfigure}[t]{0.3\textwidth}
    \centering
    \includegraphics[scale=0.2]{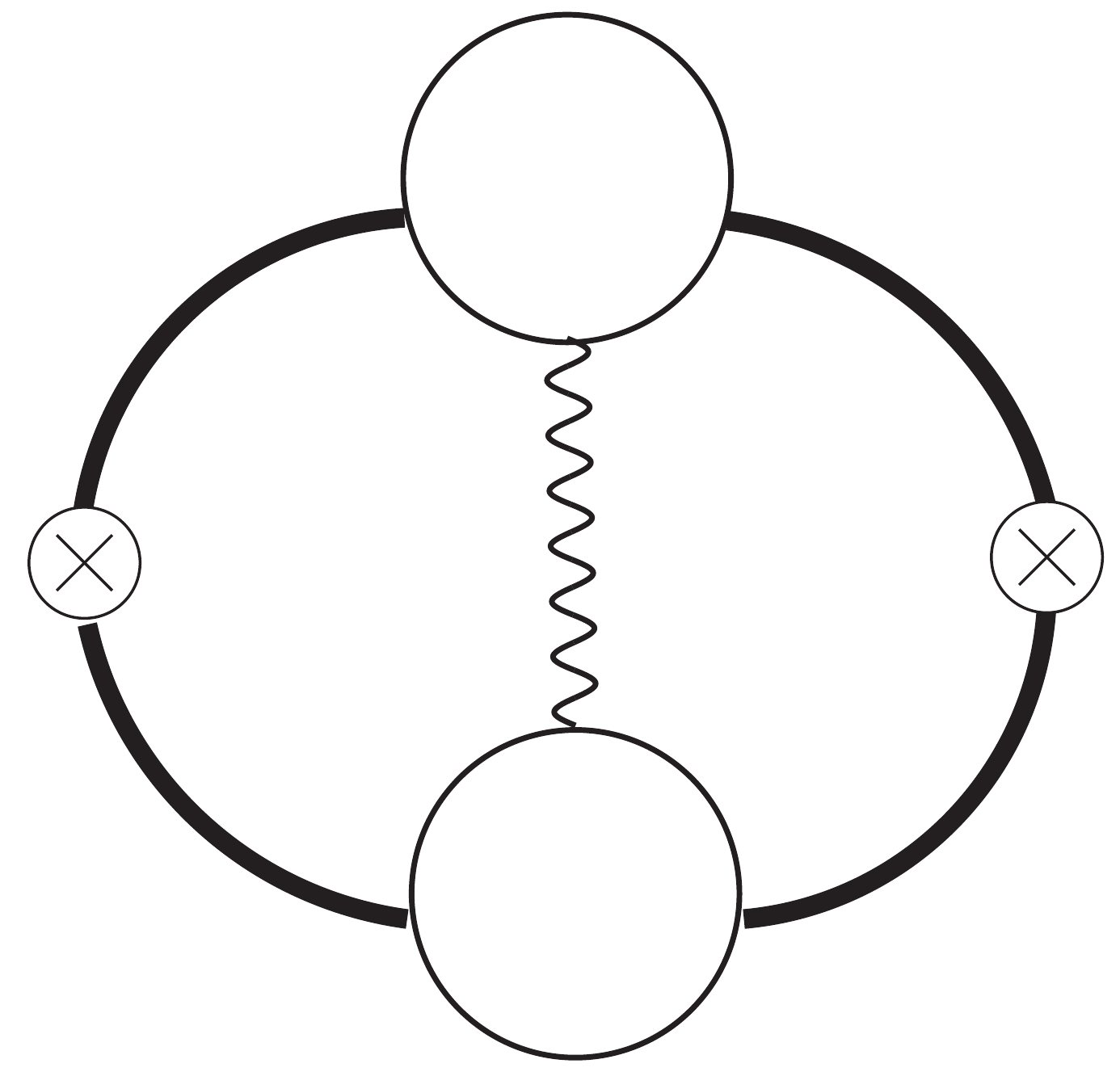}
\caption{}    
    \end{subfigure}
    \hfill
     \begin{subfigure}[t]{0.3\textwidth}
     \centering
      \includegraphics[scale=0.15]{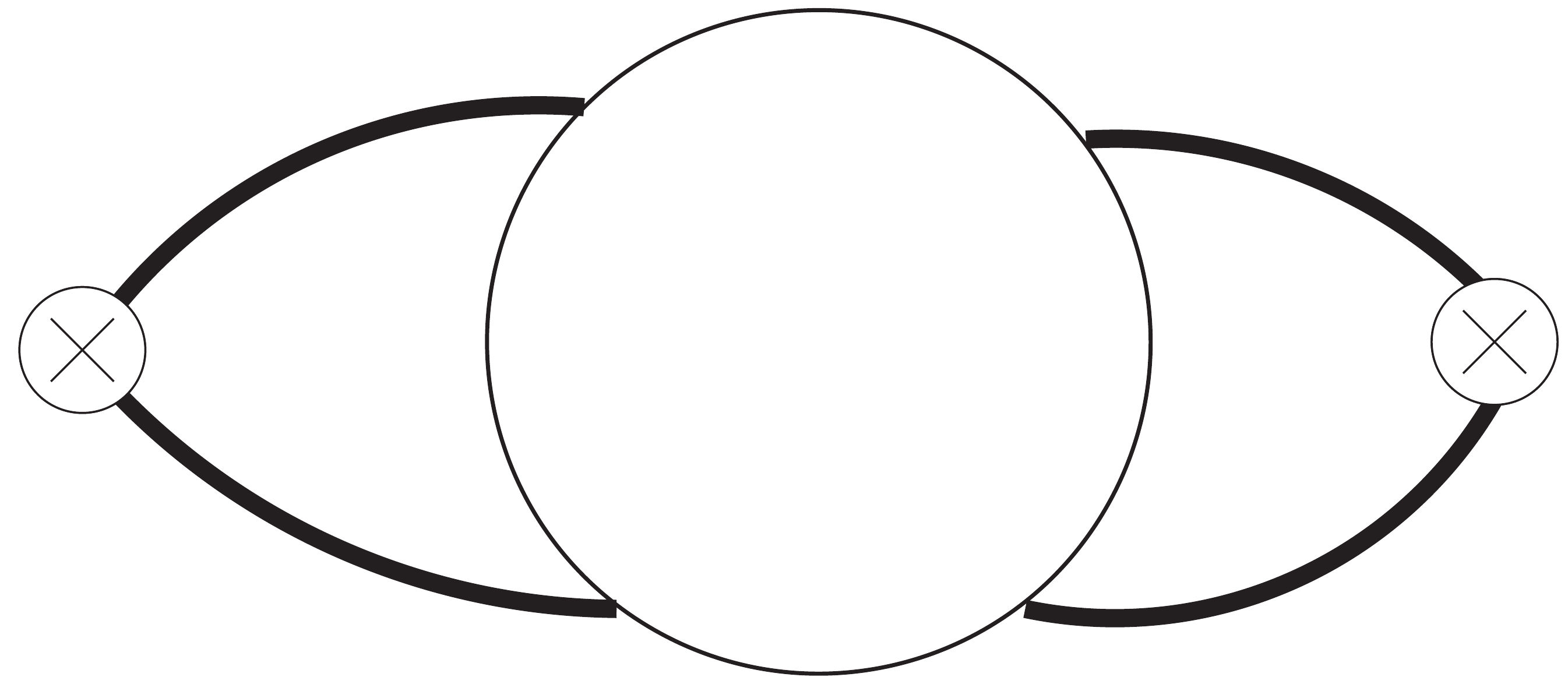}
\caption{}     
     \end{subfigure}
     \hfill
      \begin{subfigure}[t]{0.3\textwidth}
      \centering
    \includegraphics[scale=0.15]{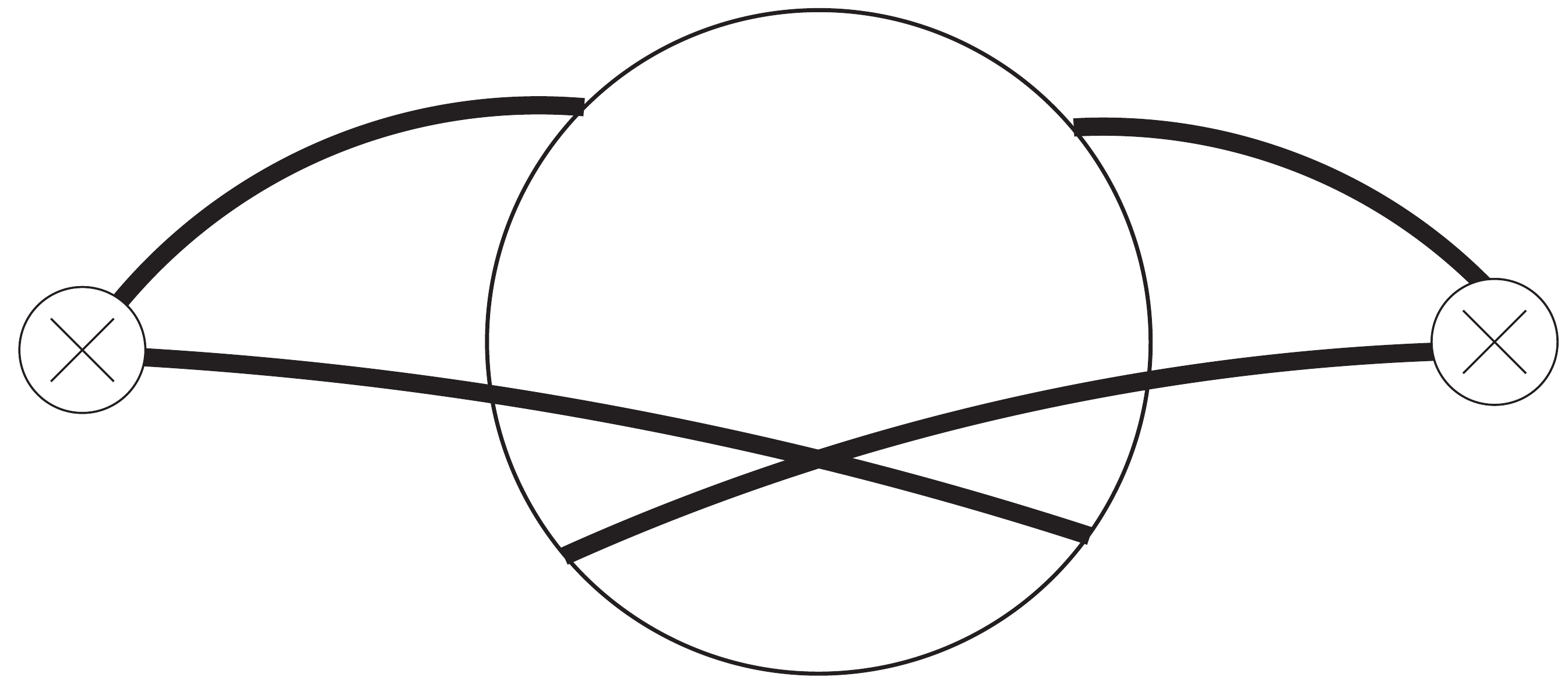}
\caption{}    
    \end{subfigure}
    \caption{Three different topologies contributing to the anomalous dimensions of the double trace scalars at the next-to-leading order in $1/N$. These diagrams are built off effective vertices in Fig.\ref{effective_vertices}. The cross-caps correspond to the insertion points of the double trace operators. The solid bold lines denote propagators of the single trace operators.}
     \label{topology-2}
 \end{figure}

\noindent
The diagrams which appear at the first sub-leading order in the $1/N$ expansion have the same topologies for all the three operators. One of them is simple. It corresponds to the sub-leading corrections of the single trace operators $\s_\pm$, and we therefore present the remaining topologies only. They are shown in Fig.\ref{topology-2}. In this section, we evaluate these diagrams for each of the double trace operators separately, by plugging in their individual propagators and combinatorial factors. 

 \subsubsection*{The $\s_+^2$ operator}
 
\begin{figure}[H]
\centering
 \begin{subfigure}[t]{1\textwidth}
    \centering
    \includegraphics[scale=0.2]{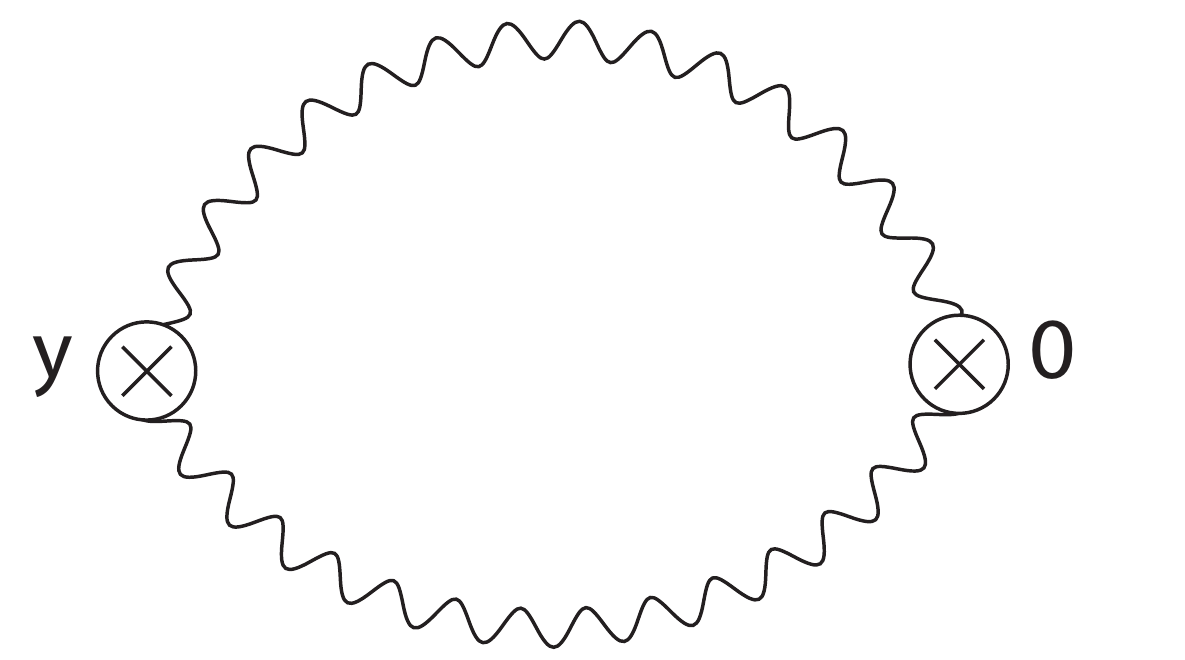}   
    \caption*{(a)} 
    \end{subfigure}
    \begin{subfigure}[t]{0.2\textwidth}
    \centering
    \includegraphics[scale=0.25]{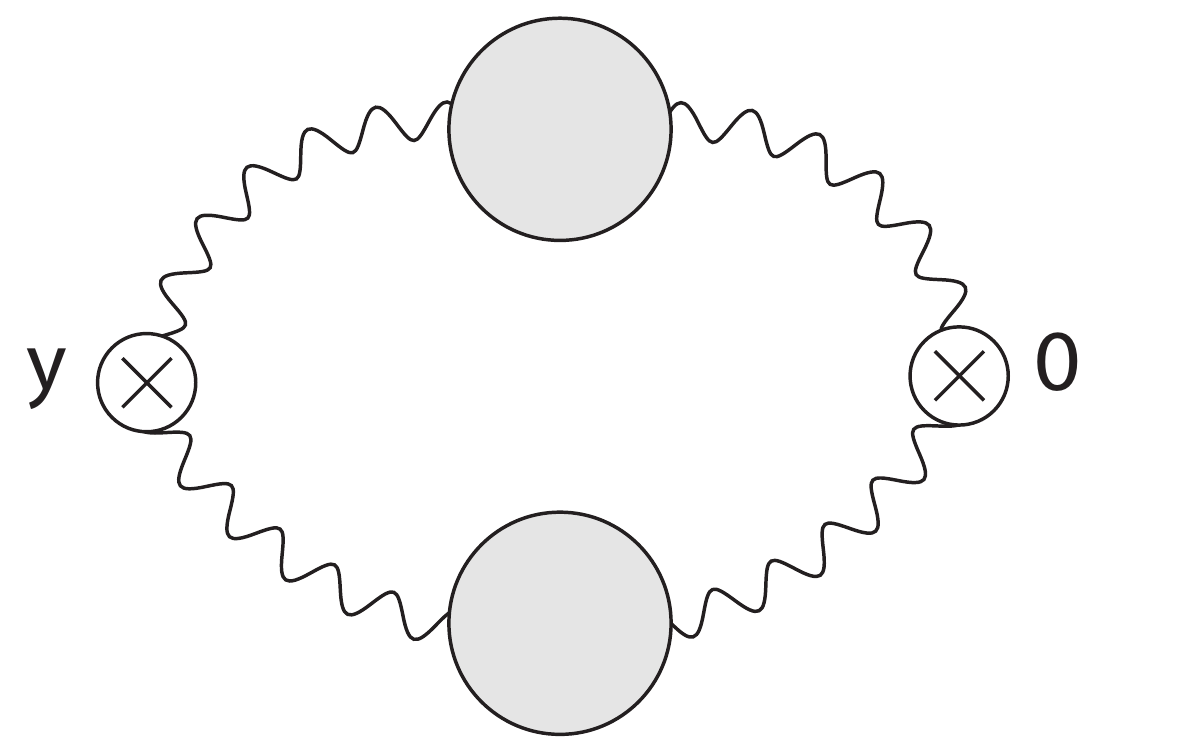}
        \caption*{(b)}
    \end{subfigure}
     \begin{subfigure}[t]{0.2\textwidth}
     \centering
      \includegraphics[scale=0.25]{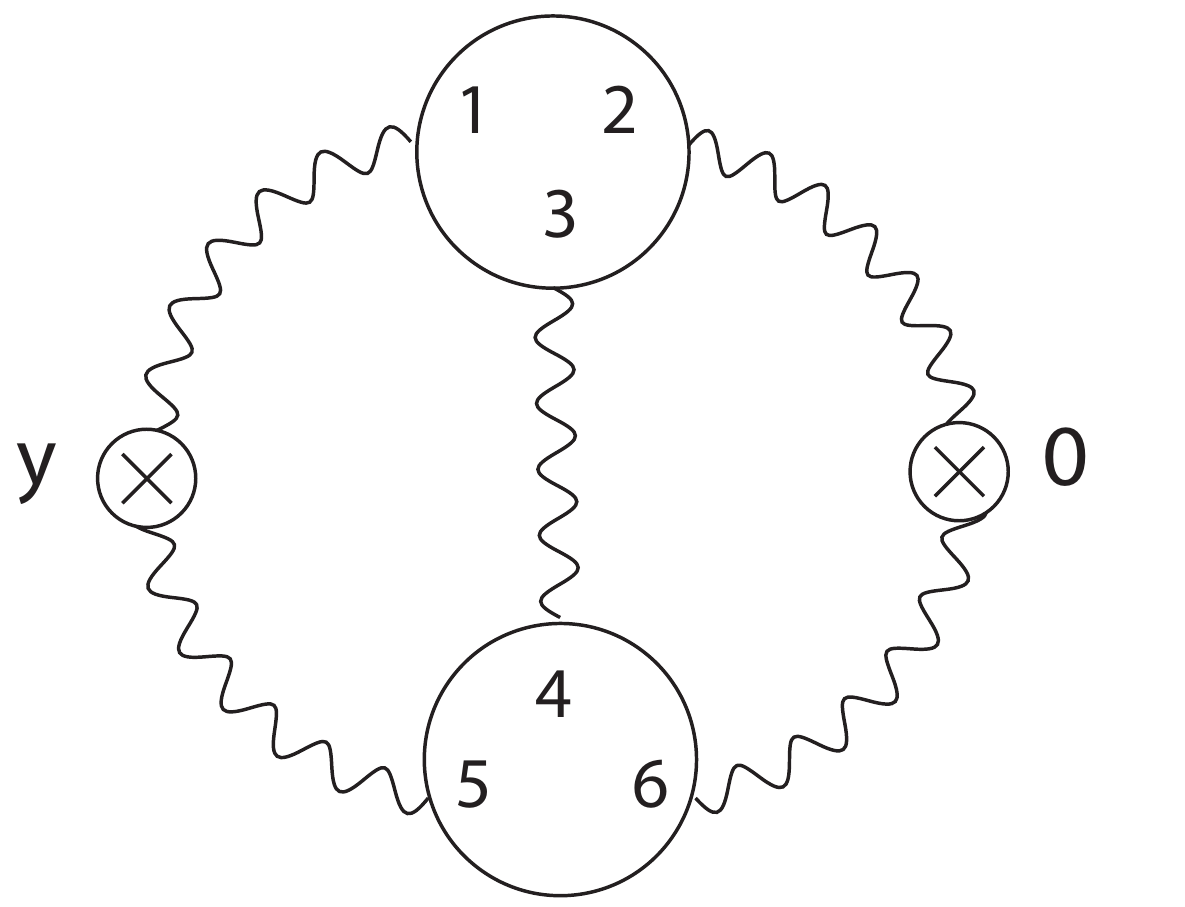}
       \caption*{(c)}
     \end{subfigure}
      \begin{subfigure}[t]{0.2\textwidth}
      \centering
    \includegraphics[scale=0.25]{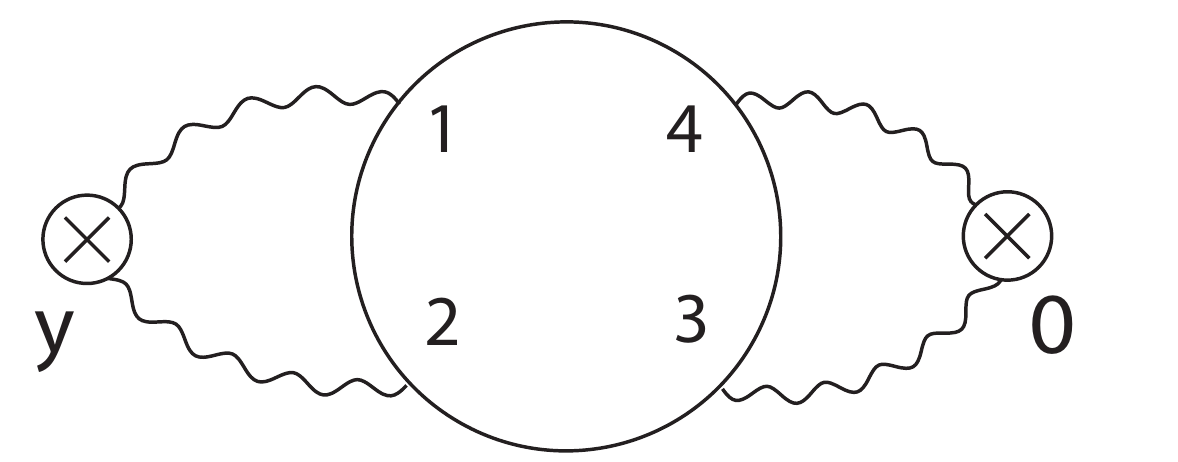}
     \caption*{(d)}
    \end{subfigure}
    \begin{subfigure}[t]{0.2\textwidth}
      \centering
    \includegraphics[scale=0.25]{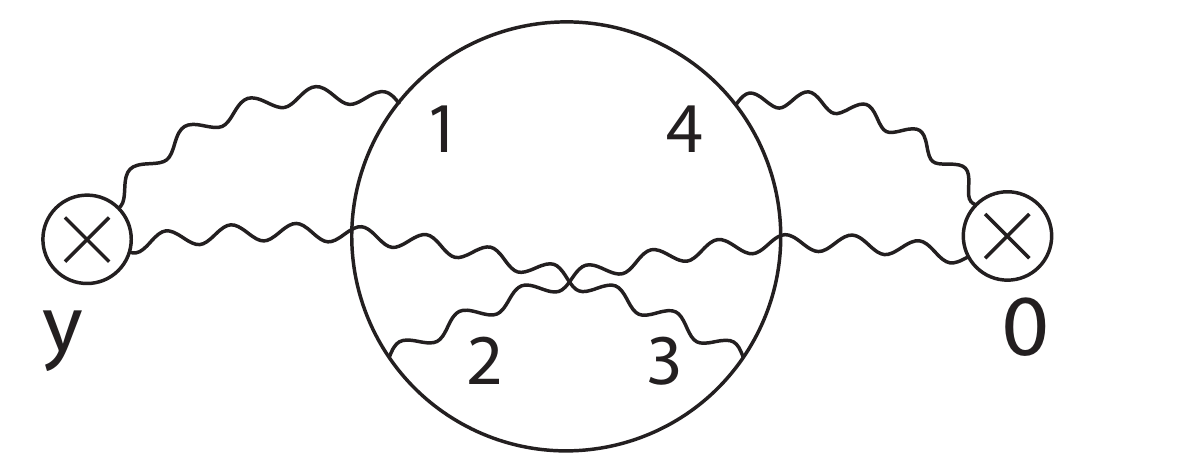}
     \caption*{(e)}
    \end{subfigure}
    \caption{Feynman diagrams contributing to the $\lan \s^2_+ \s^2_+\ran$ correlator up to $1/N$ order: (a) leading order diagram (b)-(e) next-to-leading order graphs. A wavy propagator with a grey blob in diagram (b) represents the full $\lan\sigma_+ \sigma_+ \ran$ propagator.} 
     \label{sigma_dt-1}
 \end{figure}

 \noindent
 The full correlator of $\sigma^2_{+}$ is given by
 \begin{equation}
   \lan\sigma^2_+(y) \sigma^2_+(0) \ran=\frac{2C_+^2(1+A_{++})}{|y|^{2(\Delta_{++}+\gamma_{++})}}  \mu^{-2\gamma_{++}}~,
   \quad \Delta_{++}=4~.
 \end{equation}
where $A_{++}\sim \mathcal{O}(1/N)$ is a sub-leading  correction to the amplitude, and $\gamma_{++}\sim \mathcal{O}(1/N)$ is the anomalous dimension. The large $N$ limit of this correlator is entirely fixed by a single diagram in Fig.\ref{sigma_dt-1}(a). In contrast, the $1/N$ correction is determined by the four diagrams in Fig.\ref{sigma_dt-1}(b)-(e).

 \noindent
The contribution of Fig.\ref{sigma_dt-1}(b) to $\gamma_{++}$ is simply $2\gamma_+$, and we proceed to the rest of the diagrams. The graph in Fig.\ref{sigma_dt-1}(c) is built off two effective cubic vertices in Fig.\ref{effective_vertices} with $U^i_+\sigma_+$ substituted for $\sigma_i$ and contracted with two external operators $\sigma_+^2$. The diagram itself can be calculated with the help of repeated use of the star-triangle identity \eqref{uniqueness scalar}. We relegate the details of calculation to Appendix \ref{appx:identities} (see \eqref{figpp_c} there),
\bea
 \text{Fig.}\ref{sigma_dt-1}(c)&=& {1\over N} {4(8\pi^d)^3 (d-2) \Gamma^2\({2-d\over 2}\)\over (d-4)^3 \, \Gamma^2(d-3)}
 \, {\log\mu\over |y|^8} \, 
 \sum_{i,j} {(U_+^i U_+^j)^3\over \sqrt{x_i x_j}} C_\phi^6 C_+^5 
 \nn \\
 &=& {32\over N} {(d-3)^2 \Gamma(d-1) \sin\({\pi d\over 2}\) \over \pi \Gamma^2\({d\over 2}\)}
 \,C_+^2 \, {\log\mu\over |y|^8} \, \sum_{i,j} {h_{ij}^3\over \sqrt{x_i x_j} \, \text{Tr}^3 (\bh)}  ~,
\eea
where in the second equality we used \eqref{completeness}.

\begin{figure}[t!]
\centering
   \includegraphics[scale=0.5]{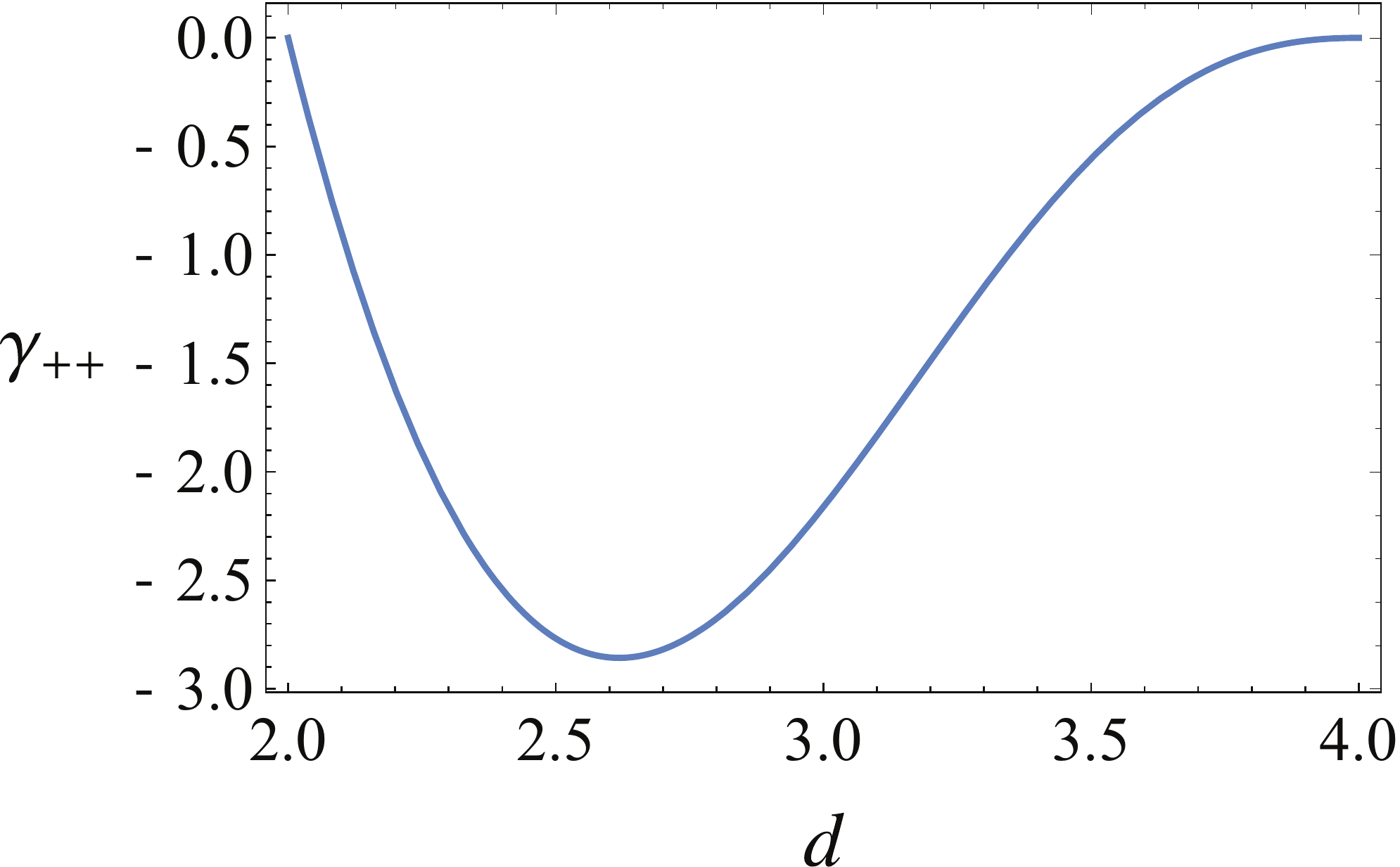}
    \caption{The anomalous dimension $\gamma_{++}$ as a function of $d$ for $h_{12}<0$. For simplicity we set $N=1$.} 
     \label{pp}
 \end{figure}

\noindent
Similarly, using the effective quartic vertex in Fig.\ref{effective_vertices}, with $U^i_+\sigma_+$ substituted for $\sigma_i$, and contracting it with the external operators $\sigma_+^2$, yields 
\bea
 \text{Fig.}\ref{sigma_dt-1}(d)&=& -{128\over N} 
 {\pi^{2d} (d-2) \Gamma\({2-d\over 2}\)\over (d-4) \, \Gamma(d-3)\Gamma(d/2)}
 \, {\log\mu\over |y|^8} \, 
 \sum_i {(U_+^i)^4\over x_i} C_\phi^4 C_+^4 
  \nn \\
 &=&     - {8\over N} {(d-3)(d-4)\Gamma(d-1) \sin\({\pi d\over 2}\) \over \pi \Gamma^2\({d\over 2}\)}
 \,C_+^2 \, 
 {\log\mu\over |y|^8} \, \sum_i {h_{ii}^2\over x_i \text{Tr}^2 (\bh)}  ~,
\eea
where in the first equality we repeatedly used the star-triangle relation \eqref{uniqueness scalar} to carry out the loop integrals, whereas in the second equality we substituted  \eqref{completeness}. Finally, based on \eqref{figpp_e}, we obtain
\bea
 \text{Fig.}\ref{sigma_dt-1}(e)&=& 
 {(32\pi^d)^2 \Gamma\({8-d\over 2}\)\over N(d-4)^3 \, \Gamma(d-3)\Gamma(d/2)}
 \, {\log\mu\over |y|^8} \, 
 \sum_i {(U_+^i)^4\over x_i} C_\phi^4 C_+^4 
  \nn \\
 &=& - {8\over N} {(d-3)(d-6)\Gamma(d-1) \sin\({\pi d\over 2}\) \over \pi \Gamma^2\({d\over 2}\)}
 \,C_+^2 \, 
 {\log\mu\over |y|^8} \, \sum_i {h_{ii}^2\over x_i \text{Tr}^2 (\bh)}  ~,
\eea 
Combining all together, yields
\be
 \gamma_{++}=2\gamma_+ + {4\over N} {(d-3)\Gamma(d-1) \sin\({\pi d\over 2}\) \over \pi \Gamma^2\({d\over 2}\)}
  \, \sum_{i,j} \( {(d-5)h_{ii}^2 \delta_{ij}\over x_i \text{Tr}^2 (\bh)}  -  {2(d-3)h_{ij}^3\over \sqrt{x_i x_j} \text{Tr}^3 (\bh)}\)
  ~.
  \label{gamma_pp}
\ee 
The dependence on the critical values of the couplings drops out for the equal rank case, $x_1=x_2=1/2$, because the $\mathbb{Z}_2$ symmetry, which swaps between $\phi_1$ and $\phi_2$, is respected by the fixed points of interest, \ie in this case $h_{11}=h_{22}=|h_{12}|$.\footnote{It can be explicitly verified using the standard $\epsilon$-expansion around $d=4$.} Hence,
 \bea
 \gamma_{++}\Big|_{m={N\over 2}}&=& {2^{d-1}(d-4)(d^2-4d-1)\Gamma\({d-1\over 2}\) \sin\({\pi d\over 2}\)\over N \pi^{3\over 2} \Gamma\({d+2\over 2}\)}
   \quad \text{for} \quad h_{12}<0 ~,
  \nn \\
   \gamma_{++}\Big|_{m={N\over 2}}&=&-{2^d (d-4)(d-1)\Gamma\({d+1\over 2}\) \sin\({\pi d\over 2}\)\over N \pi^{3\over 2} \Gamma\({d+2\over 2}\)}
      \quad \text{for} \quad h_{12}>0 ~.
      \label{eqrank gamma_pp}
\eea
The plot of $\gamma_{++}$ for negative $h_{12}$ is shown in Fig. \ref{pp}. 

\noindent
As expected, \eqref{eqrank gamma_pp} for $h_{12}>0$ matches the anomalous dimension of the Hubbard-Stratonovich field in the critical $O(N)$ model \cite{Ma:1974qh,Lang:1993ct}. Indeed, as stressed in section \ref{sec:setup}, the symmetry of the bi-conical model is enhanced to $O(N)$ provided that ranks are equal and $h_{12}>0$. Now the single trace scalar, $s\sim \phi_1^2 + \phi_2^2$,  of the enhanced symmetry corresponds to the Hubbard-Stratonovich field of the critical $O(N)$ model. In terms of auxiliary fields of the equal rank bi-conical model, it is given by $s \sim \sigma_1+\sigma_2\sim \sigma_+$, and therefore its two point function is entirely determined by the $\lan \sigma_+\sigma_+\ran$ correlator.

 \subsubsection*{The $\s_-^2$ operator}

 As in the case of $\sigma_+^2$, the anomalous dimension of $\sigma^2_{-}$ is determined by matching the diagrammatic expansion of $\lan\sigma^2_- \sigma^2_- \ran$ with the general form of the exact correlation function
 \begin{equation}
   \lan\sigma^2_-(y) \sigma^2_-(0) \ran=\frac{2C_-^2(1+A_{--})}{|x|^{2(\Delta_{--}+\gamma_{--})}}  \mu^{-2\gamma_{--}}~,
   \quad \Delta_{--}=2(d-2)~, \quad C_-=8\lambda_-^2C_\phi^2~,
 \end{equation} 
where $A_{--}\sim \mathcal{O}(1/N)$ is associated with a sub-leading  correction to the leading order amplitude, and $\gamma_{--}\sim \mathcal{O}(1/N)$ is the anomalous dimension. 

\begin{figure}[t!]
\centering
 \begin{subfigure}[t]{1\textwidth}
    \centering
    \includegraphics[scale=0.25]{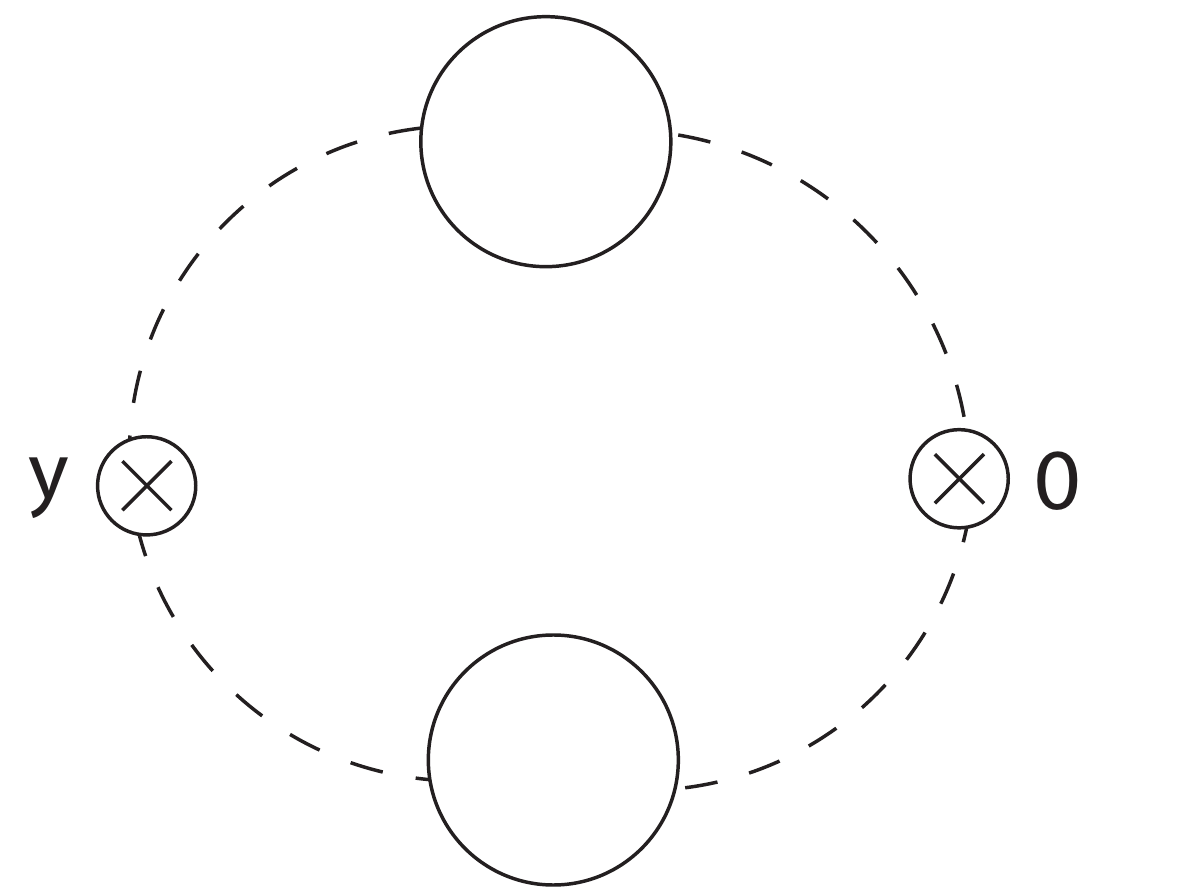}
     \caption*{(a)}
    \end{subfigure}
    \begin{subfigure}[t]{0.2\textwidth}
    \centering
    \includegraphics[scale=0.25]{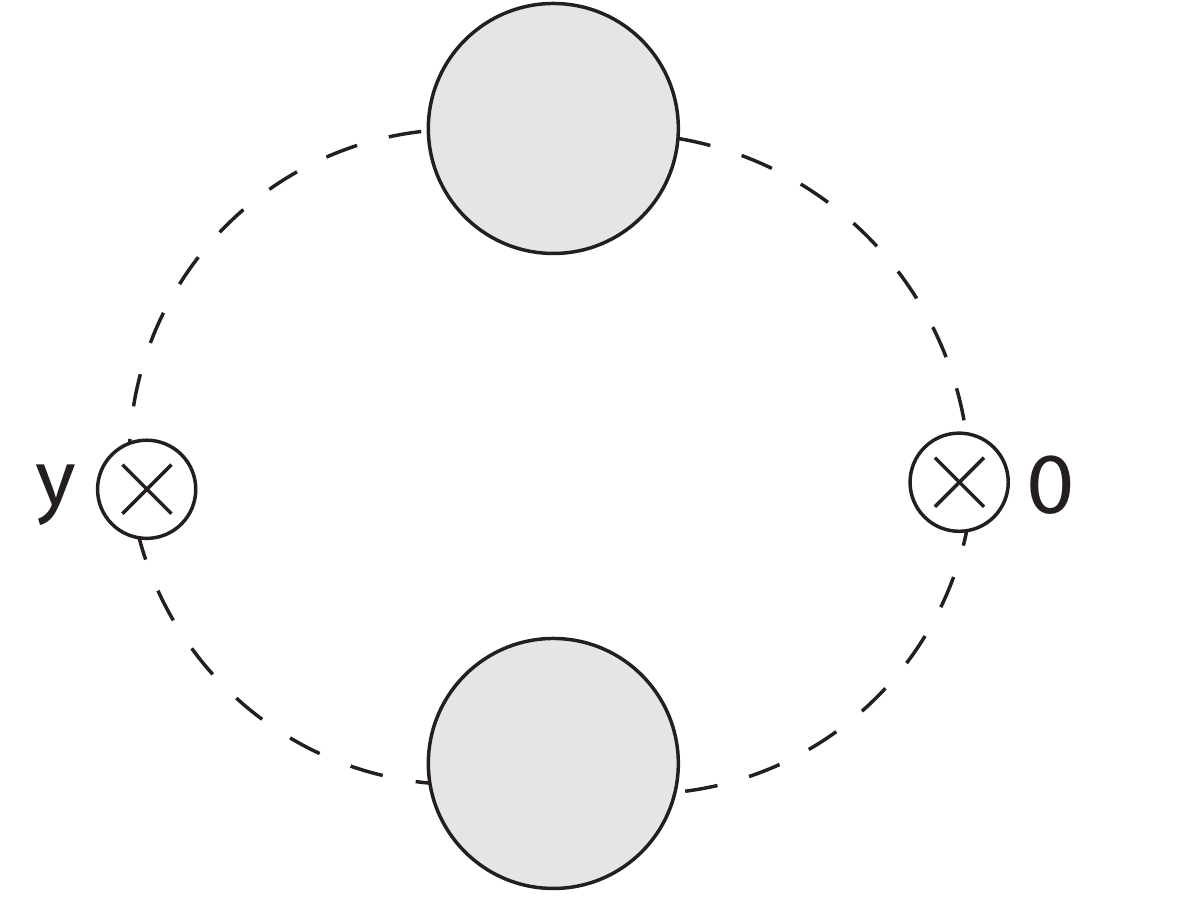}
     \caption*{(b)}
    \end{subfigure}
     \begin{subfigure}[t]{0.2\textwidth}
     \centering
      \includegraphics[scale=0.25]{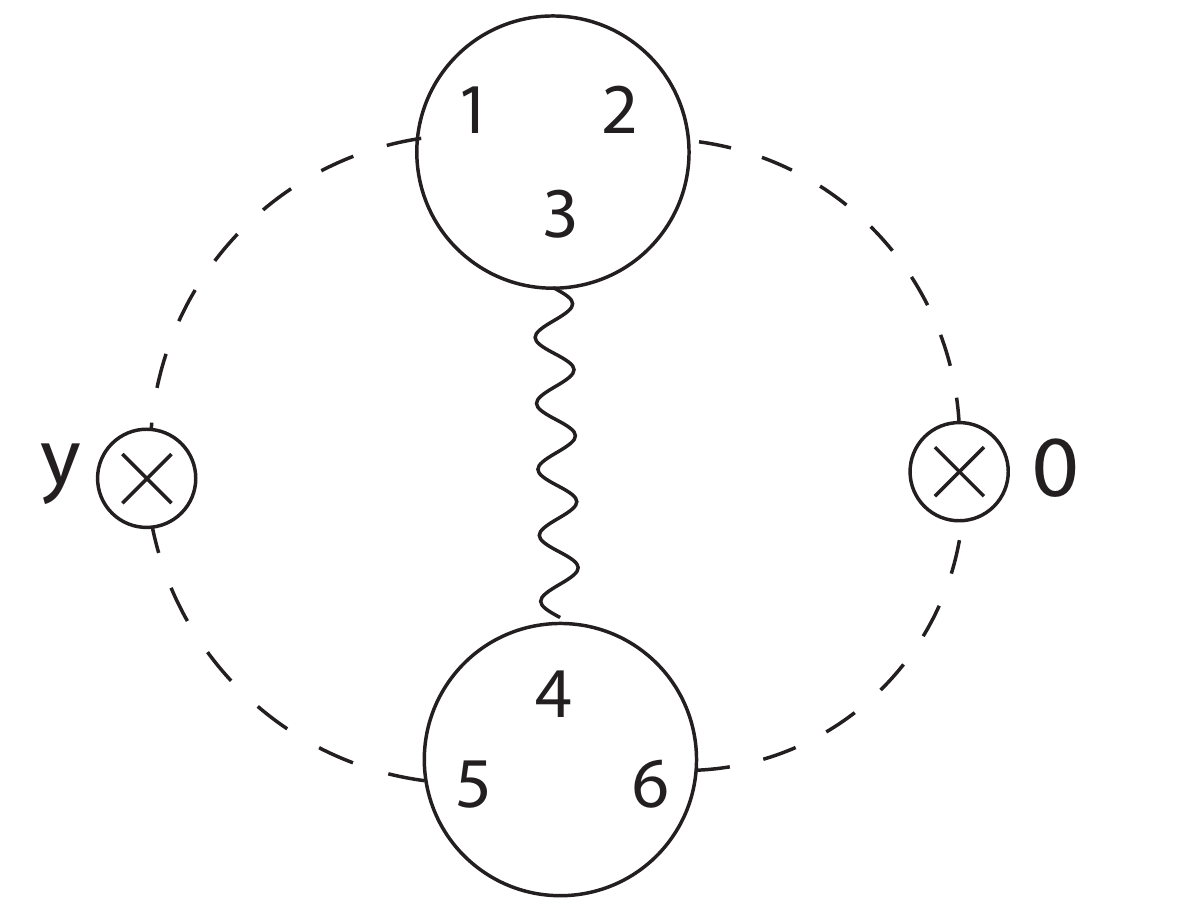}
        \caption*{(c)}
     \end{subfigure}
      \begin{subfigure}[t]{0.2\textwidth}
      \centering
    \includegraphics[scale=0.25]{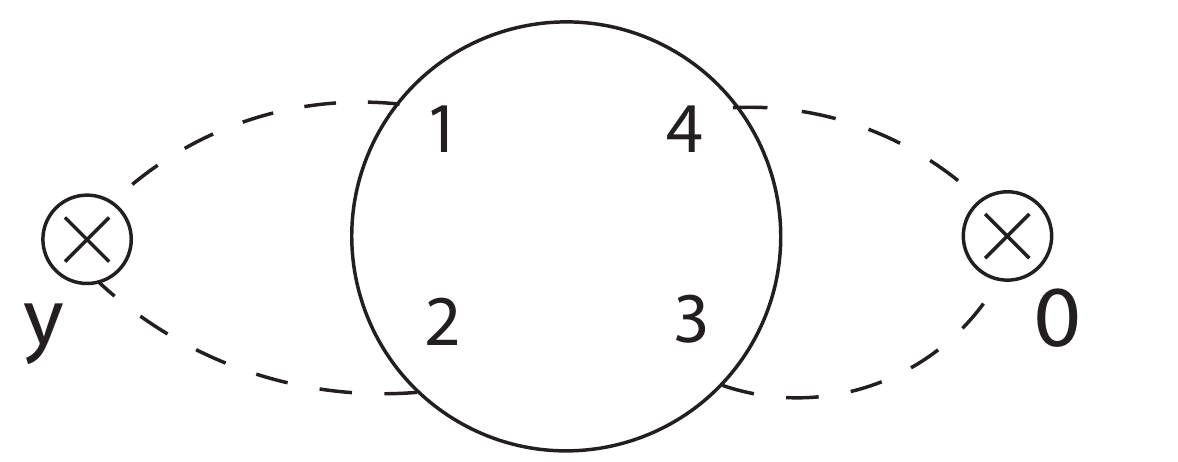}
      \caption*{(d)}
    \end{subfigure}
    \begin{subfigure}[t]{0.2\textwidth}
      \centering
    \includegraphics[scale=0.25]{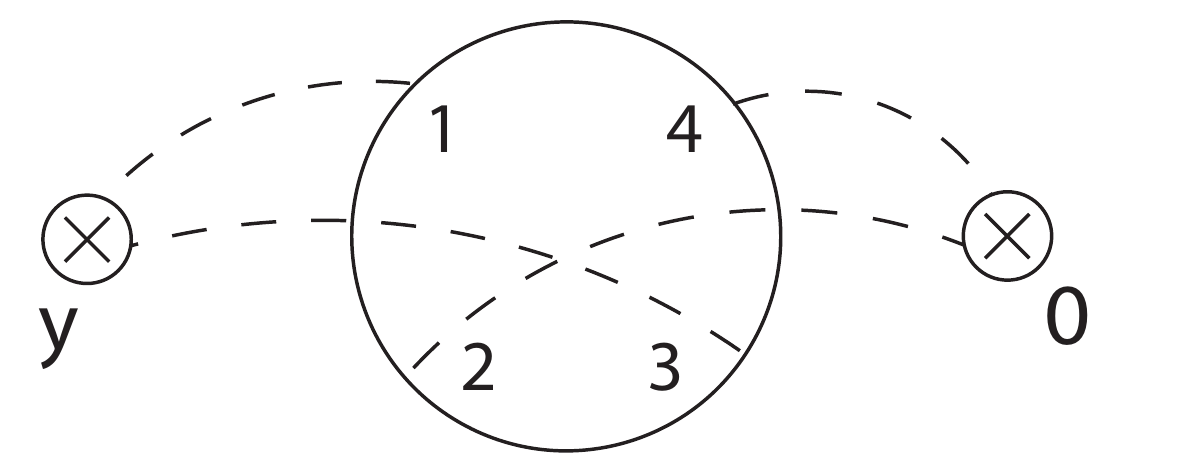}
      \caption*{(e)}
    \end{subfigure}
    \caption{Feynman diagrams contributing to the $\lan \s^2_- \s^2_-\ran$ correlator up to $1/N$ order: (a) The leading order diagram built off a sub-leading term proportional to $B(y)$ in \eqref{prop} (b)-(e) Next-to-leading order graphs. A dashed propagator with a grey blob in diagram (b) represents the full $\lan\sigma_- \sigma_- \ran$ propagator.} 
     \label{sigma_dt-2}
 \end{figure}

\noindent
The leading order behaviour of this correlator is entirely fixed by a single diagram in Fig.\ref{sigma_dt-2}(a). Note that the correlation function of the $\s_-^2$ operator, likewise that of $\s_-$, are suppressed by powers of $1/N$ relative to the analogous correlators for $\sigma_+$. This is because the dashed propagator is proportional to $\l_-\sim\mathcal{O}(1/N)$.

\noindent
The $1/N$ correction to the leading order diagram is represented in terms of four diagrams in Fig.\ref{sigma_dt-2}(b)-(e).
The diagram in Fig.\ref{sigma_dt-2}(b) contributes $2\gamma_-$ to the anomalous dimension of $\sigma_-^2$, whereas the graph of Fig.\ref{sigma_dt-2}(d) vanishes, because delta function propagators of the field $\sigma_-$ result in a term proportional to $\lan\phi_i^2\ran^2$, which vanishes in a CFT. Similarly, the diagram in Fig.\ref{sigma_dt-2}(e) is finite. It contributes to $A_{--}$ only, and we ignore it in what follows.

\begin{figure}[t!]
\centering
   \includegraphics[scale=0.5]{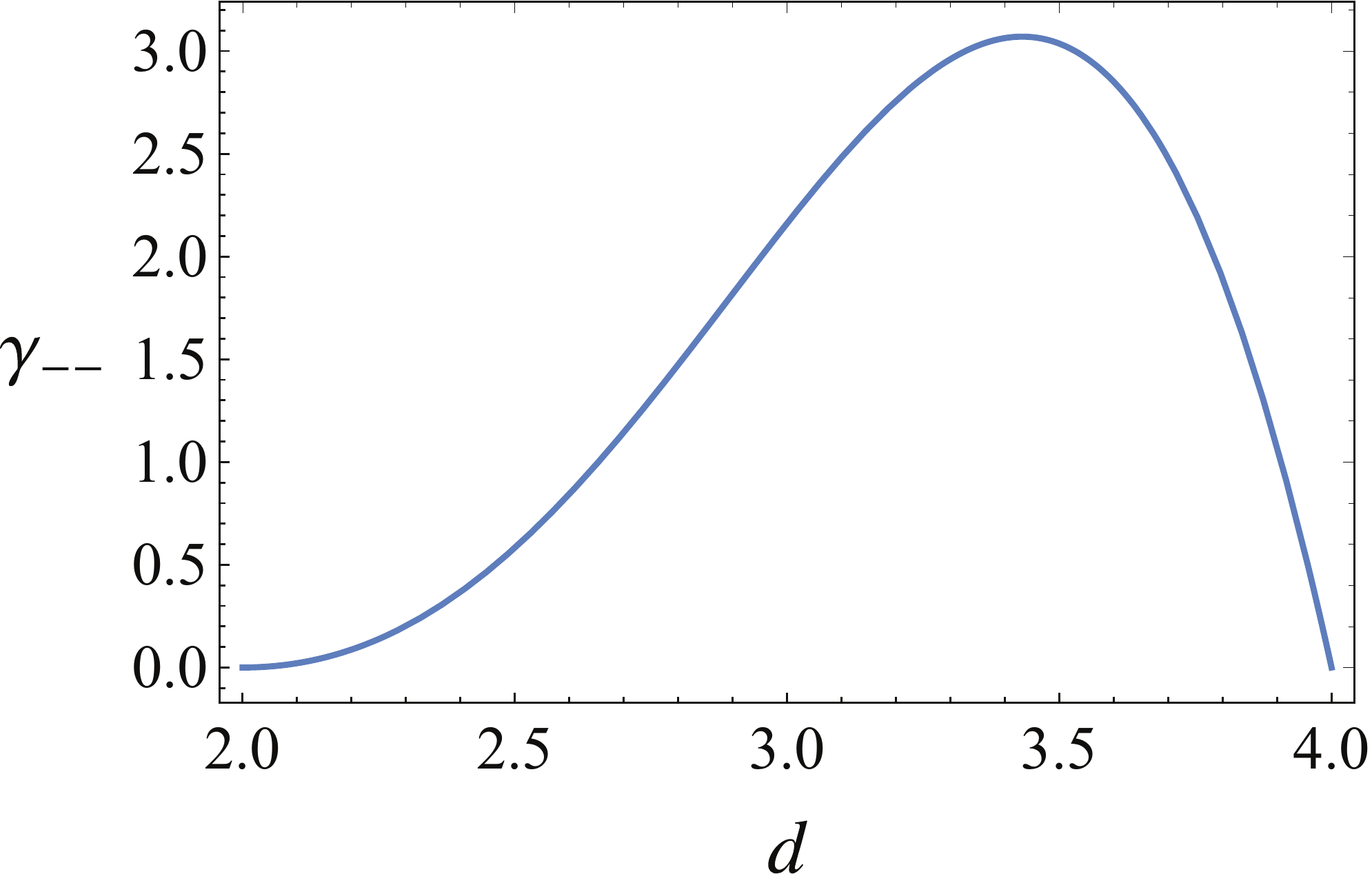}
    \caption{The anomalous dimension $\gamma_{--}$ as a function of $d$ for $h_{12}<0$. For simplicity we set $N=1$.} 
     \label{mm}
 \end{figure}

\noindent
The upshot of this discussion is that we have only to calculate the graph in Fig.\ref{sigma_dt-2}(c). It is built off two effective cubic vertices in Fig.\ref{effective_vertices} with $U^i_+\sigma_+$ substituted for the end points of the internal propagator of $\sigma_i$ (wavy line), and $U^i_-\sigma_-$ replacing $\sigma_i$'s of the cubic vertices which are contracted with two external operators $\sigma_-^2$. The diagram itself can be calculated with the help of repeated use of the star-triangle identity \eqref{uniqueness scalar} (see \eqref{figmm_c} in Appendix \ref{appx:identities} for details),  
\bea
 \text{Fig.}\ref{sigma_dt-2}(c)&=& 
 {512\over N} {(d-2) \pi^d\over (d-4)\Gamma^2\({d\over 2}\)}\, {\log\mu\over |y|^{4(d-2)}} \, 
 \sum_{i,j} {(U_-^i U_-^j)^2 U_+^i U_+^j\over \sqrt{x_i x_j}} C_\phi^6 C_+  (-2\lambda_-)^4 
  \\
 &=&  {32 \Gamma^2\({d-2\over 2}\)\Gamma(d-2)\sin\({\pi d\over 2}\)\over \pi^{2d+1}(d-2)}
 \, {\lambda_-^4\over N} \, {\log\mu\over |y|^{4(d-2)}} \,  \sum_{i,j} {h_{ij} \over \sqrt{x_i x_j} \, \text{Tr} (\bh) }
 \( \delta_{ij} - {h_{ij} \over \text{Tr} (\bh) } \)^2 \,,
 \nn
\eea
where in the second equality we used \eqref{completeness}. Combining all together, yields
\be
 \gamma_{--}=2\gamma_-  
 -{8 \, \Gamma(d-1)\sin\({\pi d\over 2}\)\over N \pi \Gamma^2\({d\over 2}\) }
  \sum_{i,j} {h_{ij} \over \sqrt{x_i x_j} \, \text{Tr} (\bh) }
 \( \delta_{ij} - {h_{ij} \over \text{Tr} (\bh) } \)^2 ~.
 \label{gamma_mm}
\ee 
In the case of equal rank $x_1=x_2=1/2$, this expression simplifies
 \bea
 \gamma_{--}\Big|_{m={N\over 2}}&=& -{2^{d+1} (d-2)\Gamma\({d+1\over 2}\) \sin\({\pi d\over 2}\)\over N \pi^{3\over 2} \Gamma\({d+2\over 2}\)}
   \quad \text{for} \quad h_{12}<0 ~,
  \nn \\
   \gamma_{--}\Big|_{m={N\over 2}}&=&-{2^{d+1} (d+2)\Gamma\({d+1\over 2}\) \sin\({\pi d\over 2}\)\over N (d-1)\pi^{3\over 2} \Gamma\({d+2\over 2}\)}
      \quad \text{for} \quad h_{12}>0 ~.
      \label{eqrank gamma_mm}
\eea
The plot of $\gamma_{--}$ for negative $h_{12}$ is shown in Fig. \ref{mm}. 

\subsubsection*{The $\s_+\s_-$ operator}

The operator $\s_+\s_-$ is marginal to leading order in the large-$N$ expansion. Hence, in this limit there exists a conformal manifold in general $d$. Calculating the anomalous dimension at the next-to-leading order in $1/N$ will tell us whether the model exhibits a conformal manifold at finite $N$. If the anomalous dimension does not vanish, it means that this manifold is lifted by the $1/N$ corrections.

\begin{figure}[H]
\centering
 \begin{subfigure}[t]{1\textwidth}
    \centering
    \includegraphics[scale=0.25]{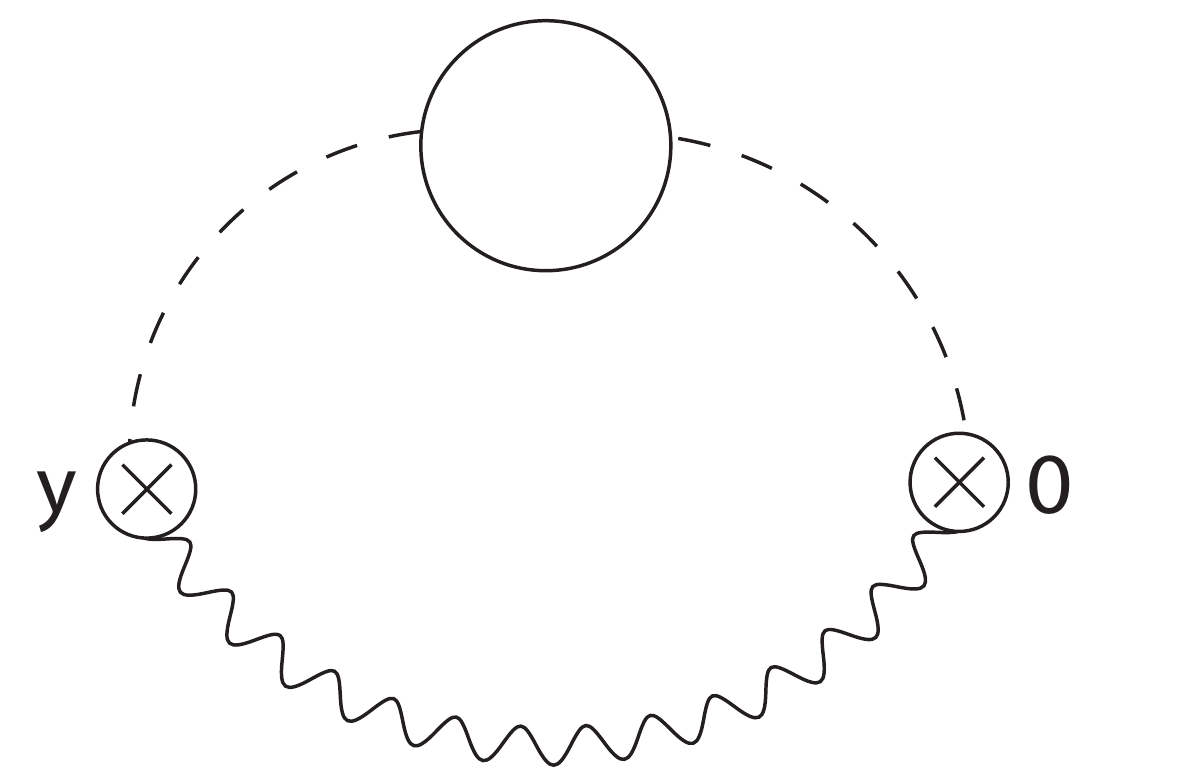}
  \caption*{(a)}    
    \end{subfigure}
    \begin{subfigure}[t]{0.2\textwidth}
    \centering
    \includegraphics[scale=0.25]{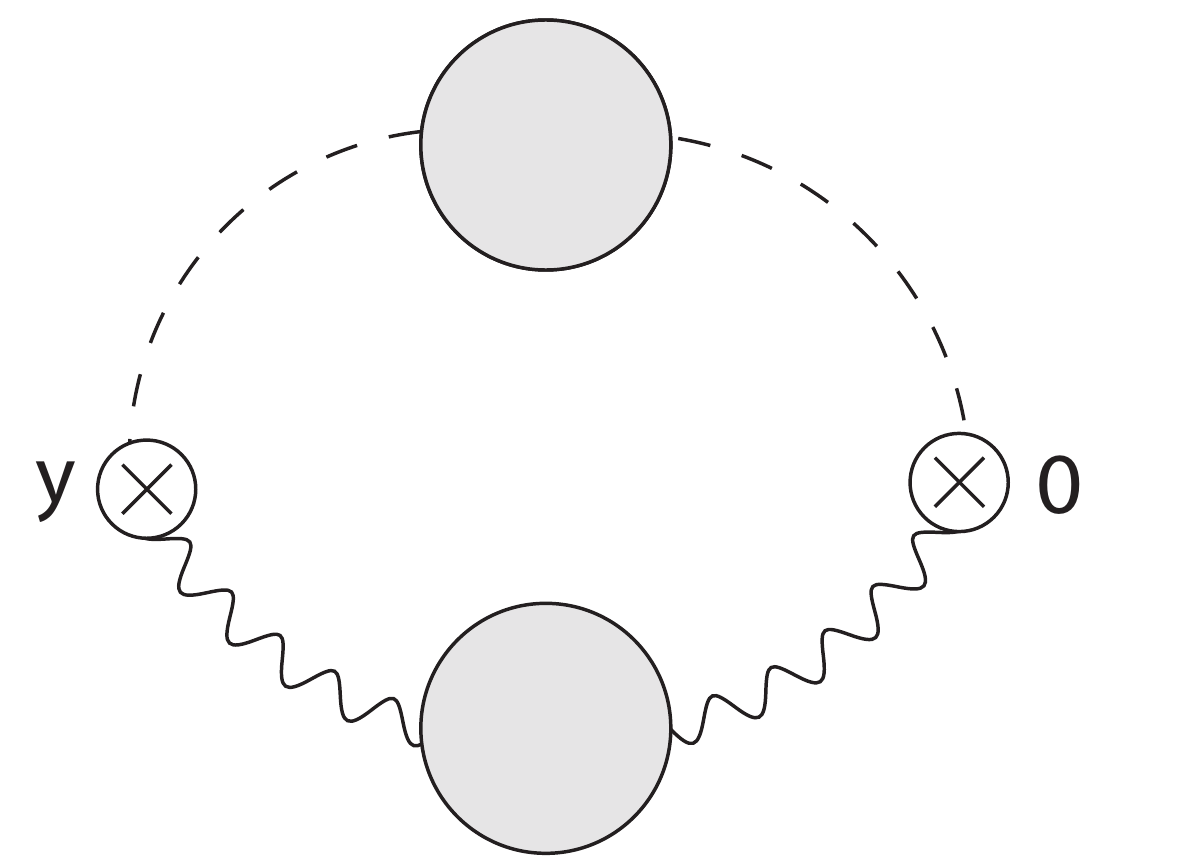}
      \caption*{(b)}
    \end{subfigure}
      \begin{subfigure}[t]{0.2\textwidth}
      \centering
    \includegraphics[scale=0.25]{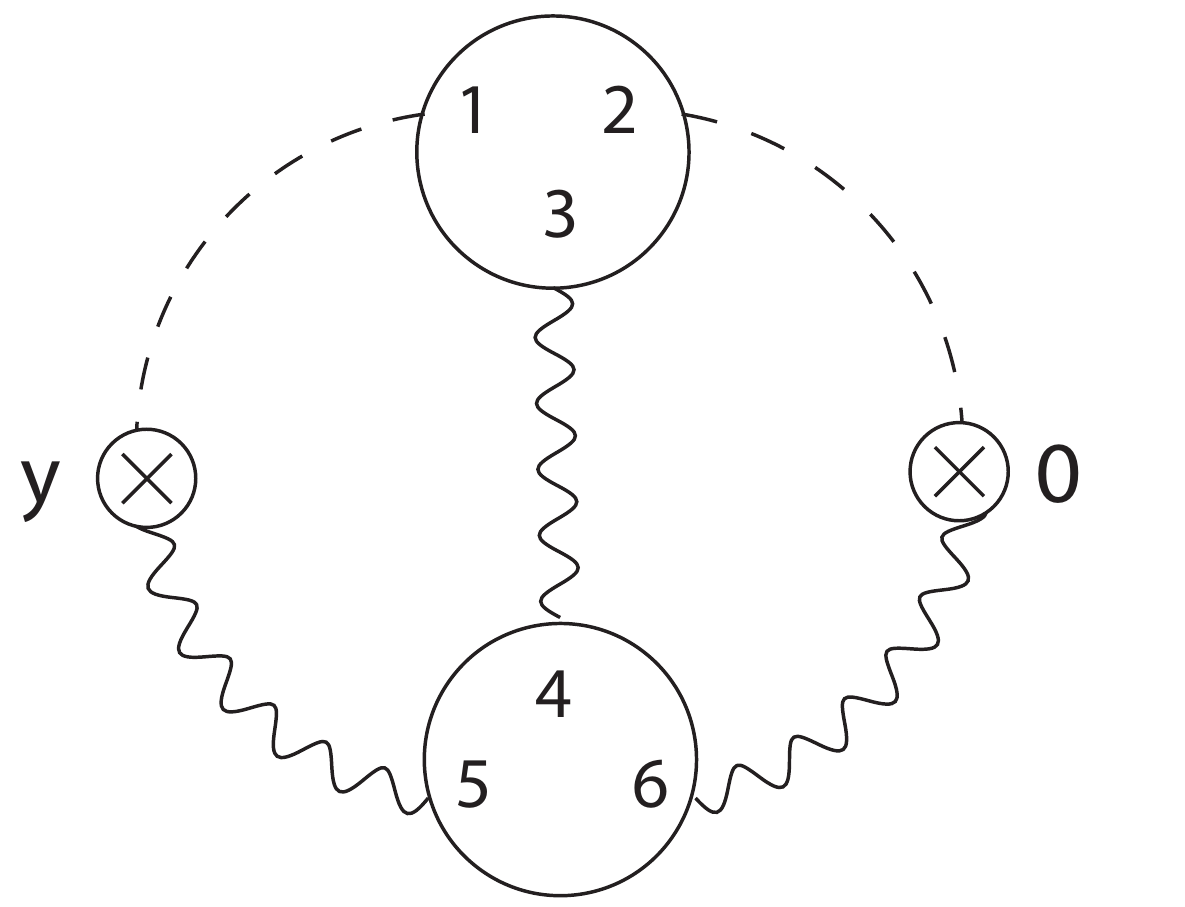}
      \caption*{(c)}
    \end{subfigure}
    \begin{subfigure}[t]{0.2\textwidth}
      \centering
    \includegraphics[scale=0.25]{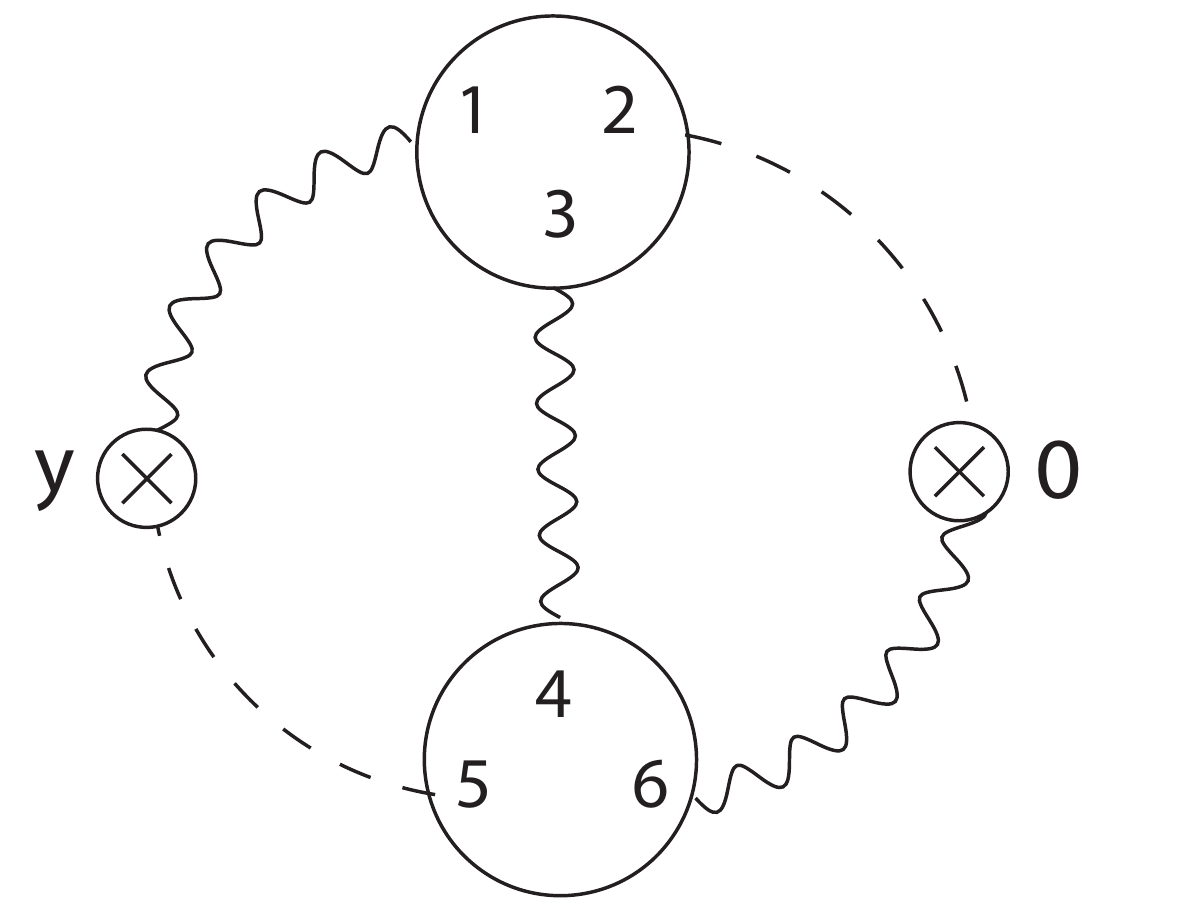}
      \caption*{(d)}
    \end{subfigure}
     \begin{subfigure}[t]{0.2\textwidth}
      \centering
    \includegraphics[scale=0.25]{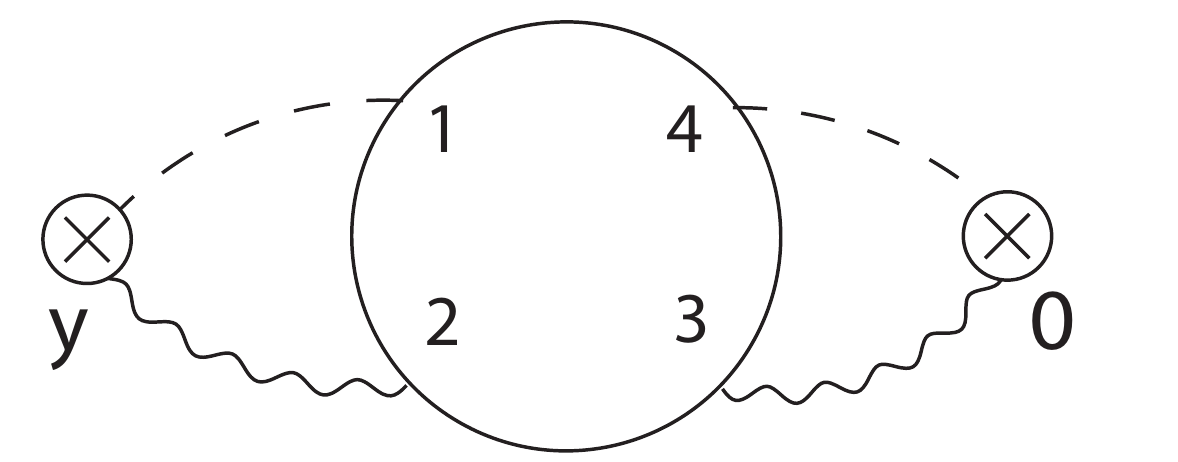}
      \caption*{(e)}
    \end{subfigure}
     \begin{subfigure}[t]{0.2\textwidth}
      \centering
    \includegraphics[scale=0.25]{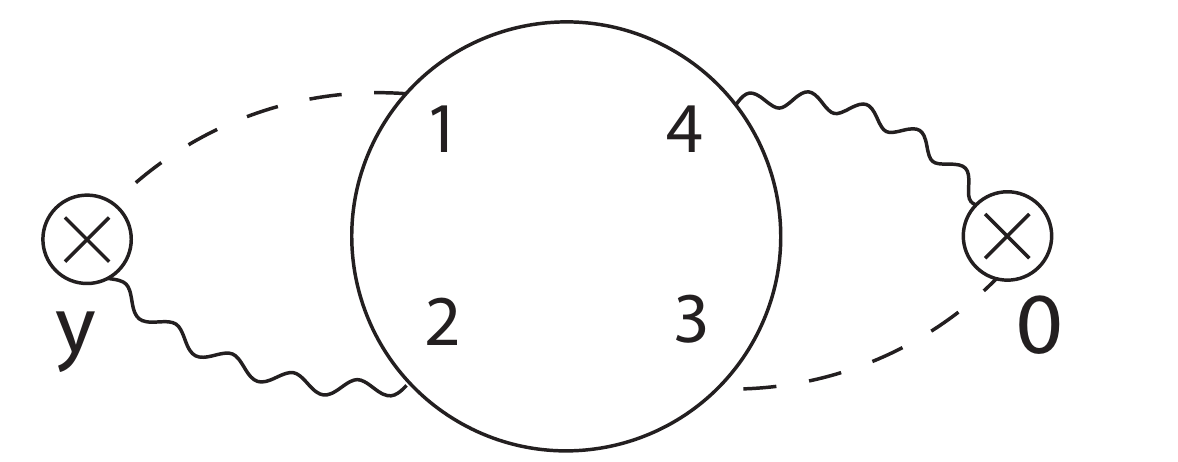}
      \caption*{(f)}
    \end{subfigure}
     \begin{subfigure}[t]{0.2\textwidth}
      \centering
    \includegraphics[scale=0.25]{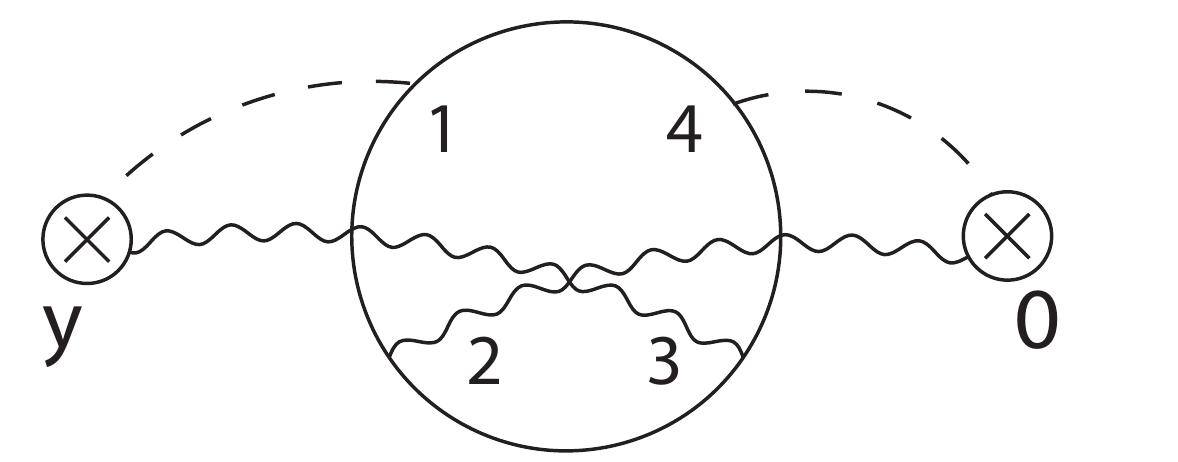}
      \caption*{(g)}
    \end{subfigure}
    \caption{Feynman diagrams contributing to \eqref{fullmarginal} up to $1/N$ order: (a) The leading order diagram built off a sub-leading term proportional to $B(y)$ in \eqref{prop} (b)-(g) Next-to-leading order graphs. The propagators with a grey blob in diagram (b) represent the full propagators of $\sigma_\pm$.}
     \label{sigma_dt-3}
 \end{figure}

\noindent
The full conformal correlator takes the form
 \begin{equation}
   \lan\s_+\s_-(y) \s_+\s_-(0) \ran=\frac{C_{+} C_-(1+A_{+-})}{|x|^{2(\Delta_{+-}+\gamma_{+-})}}  \mu^{-2\gamma_{+-}}~,
   \quad \Delta_{+-}=d~, 
   \label{fullmarginal}
 \end{equation} 
where $A_{+-}, \gamma_{+-}\sim \mathcal{O}(1/N)$ are the sub-leading  correction to the amplitude and anomalous dimension respectively. Fig.\ref{sigma_dt-3}(a) represents the leading order behaviour of this correlator. It scales as $\mathcal{O}(1/N^2)$, because the dashed propagator is proportional to $\l_-\sim\mathcal{O}(1/N)$. The $1/N$ correction  is represented in terms of six diagrams displayed in Fig.\ref{sigma_dt-3}(b)-(g).

\noindent
The contribution of Fig.\ref{sigma_dt-3}(b) to the anomalous dimension of $\sigma_-\sigma_+$ boils down to $\gamma_- + \gamma_+$. Diagrams in Fig.\ref{sigma_dt-3}(c),(d) are very similar. They are built off two effective cubic vertices in Fig.\ref{effective_vertices} with $U^i_\pm\sigma_\pm$ substituted for $\sigma_i$ in an appropriate way. Repeated use of the star-triangle identity \eqref{uniqueness scalar} eventually yields  (see \eqref{figpm_c}), 
\bea
\text{Fig.}\ref{sigma_dt-3}\text{(c)}&=& {- (32\pi^d)^2 \, \Gamma\(2-{d\over 2}\)\over N (d-4)^2\Gamma(d-3)\Gamma\({d\over 2}\)}
{\log\mu\over |y|^{2d}}\sum_{i,j} {(U_+^i)^3U_+^j (U_-^j)^2\over \sqrt{x_i x_j}} C_\phi^6 C_+^3  (-2\lambda_-)^2 
 \\
&=& {16(d-3) \Gamma(d-1)\sin\({\pi d\over 2}\) \over \pi \Gamma^2\({d\over 2}\) } C_+ C_- {\log\mu\over |y|^{2d}} 
\sum_{i,j} {h_{ii} h_{ij}\over \sqrt{x_i x_j} \, \text{Tr}^2 (\bh) } \( 1 - {h_{jj} \over \text{Tr} (\bh) } \)   ~,
\nn \\
\text{Fig.}\ref{sigma_dt-3}\text{(d)}&=& 
{16(d-3) \Gamma(d-1)\sin\({\pi d\over 2}\) \over \pi \Gamma^2\({d\over 2}\) } C_+ C_-
{\log\mu\over |y|^{2d}}
\sum_{i,j} { h_{ii} h_{jj}\over \sqrt{x_i x_j} \, \text{Tr}^2 (\bh) } \( \delta_{ij} - {h_{ij} \over \text{Tr} (\bh) } \)~.
\nn
\eea
where in the second equality we used \eqref{completeness}.

\begin{figure}[t!]
\centering
   \includegraphics[scale=0.5]{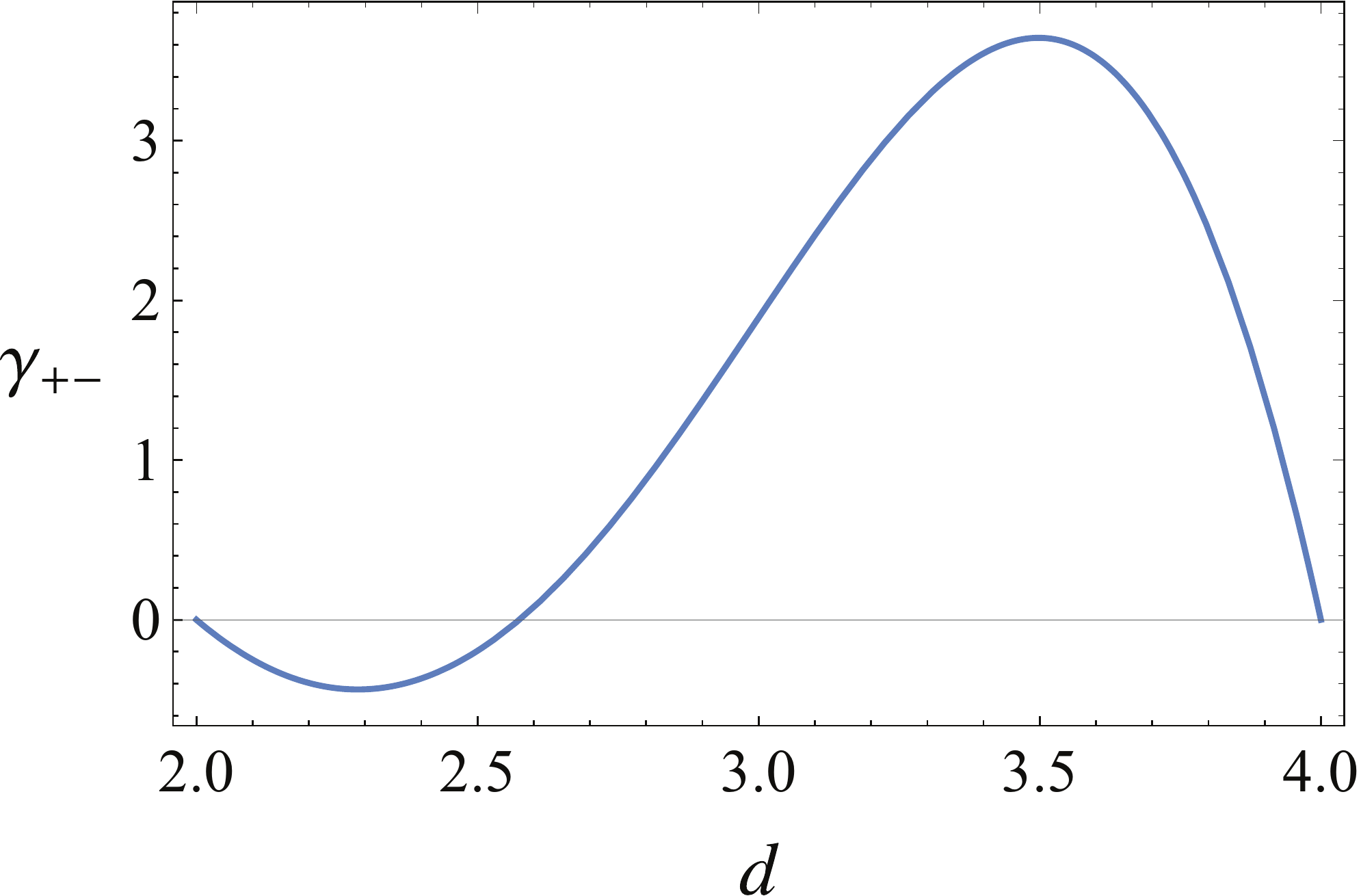}
    \caption{The anomalous dimension $\gamma_{+-}$ as a function of $d$ for $h_{12}<0$. For simplicity we set $N=1$. The sign change happens around $d=2.57437$.} 
     \label{pm}
 \end{figure} 

\noindent
Next using the identities \eqref{useful2} one can verify that the diagram in Fig.\ref{sigma_dt-3}(f) is finite, and therefore it only contributes to $A_{+-}$. In contrast,
\bea
  \text{Fig.}\ref{sigma_dt-3}\text{(e)}&=& -{32(d-2) \pi^d\over N \,d \, \Gamma^2\({d\over 2}\)}  {\log\mu\over |y|^{2d}}\sum_i {(U_+^i U_-^i)^2\over x_i}  C_\phi^4 C_+^2 (-2\lambda_-)^2
  \nn\\
  &=& - {8 \Gamma(d-2)\sin\({\pi d\over 2}\) \over  \pi \Gamma\({d-4\over 2}\) \Gamma\({d+2\over 2}\)}C_+ C_- {\log\mu\over |y|^{2d}} 
  \sum_i {h_{ii} \over x_i \text{Tr} (\bh) } \( 1 - {h_{ii} \over \text{Tr} (\bh) } \) , 
  \nn \\
  \text{Fig.}\ref{sigma_dt-3}\text{(g)}&=& {64(d-2) \pi^d\over N \,(d-4) \, \Gamma^2\({d\over 2}\)} {\log\mu\over |y|^{2d}}\sum_i {(U_+^i U_-^i)^2\over x_i}  C_\phi^4 C_+^2(-2\lambda_-)^2
   \nn\\
  &=&  {2^{d+1}\, \Gamma\({d-1\over 2}\)\sin\({\pi d\over 2}\) \over \pi^{3\over 2} \Gamma\({d\over 2}\)}
  C_+ C_- {\log\mu\over |y|^{2d}}
  \sum_i {h_{ii} \over x_i \text{Tr} (\bh) } \( 1 - {h_{ii} \over \text{Tr} (\bh) } \) ,
\eea
where we used \eqref{completeness} and \eqref{figpm_g} to get the expression for \text{Fig.}\ref{sigma_dt-3}\text{(g)}, whereas the expression for \text{Fig.}\ref{sigma_dt-3}\text{(e)} is derived based on \eqref{useful2}. As a result, we get
\bea
 \gamma_{+-}&=&{4(d-3) \Gamma(d-1)\sin\({\pi d\over 2}\) \over N \pi \Gamma^2\({d\over 2}\) } 
 \Bigg(  \, \text{sign}(h_{12})
  - 2\sum_{i,j} {h_{ii} h_{ij}\over \sqrt{x_i x_j} \, \text{Tr}^2 (\bh) } \( 1 - 2 {h_{jj} \over \text{Tr} (\bh) } \)
\nn \\
 &-& 2 \sum_i {h_{ii}^2 \over x_i \, \text{Tr}^2 (\bh) }- {(d+4)\over 2d (d-3)}  \sum_i {h_{ii} \over x_i \text{Tr} (\bh) } \( 1 - {h_{ii} \over \text{Tr} (\bh) } \) \Bigg)~.
\label{gamma_pm}
\eea
where the first term on the right hand side represents $\gamma_++\gamma_-$ given by \eqref{sing_anom}. For equal rank, we have
\bea
 \gamma_{+-}\Big|_{m={N\over 2}}&=&-
 {2(6d^2-17d+4) \Gamma(d-1)\sin\({\pi d\over 2}\) \over N \pi d\,\Gamma^2\({d\over 2}\) } 
 \quad \text{for} \quad h_{12}<0 ~,
 \nn \\
  \gamma_{+-}\Big|_{m={N\over 2}}&=&-
 {2(2d^2-5d+4) \Gamma(d-1)\sin\({\pi d\over 2}\) \over N \pi d\,\Gamma^2\({d\over 2}\) } 
 \quad \text{for} \quad h_{12}>0 ~.
 \label{gamma_pm_eqrank}
\eea 
The plot of $\gamma_{+-}$ for negative $h_{12}$ is shown in Fig. \ref{pm}. In general, $\gamma_{+-}\neq 0$ which implies that conformal manifold is lifted. However, there is a sign flip around $d_*=2.57437$, and therefore the $\sigma_+\sigma_-$ operator remains marginal in $d_*$ dimensions. In this special case, the line of fixed points might survive up to $1/N$ order. Furthermore, the higher order corrections in $1/N$ only slightly modify the value of $d_*$, because $N$ is large by assumption. Note that if the value  of $N$ is decreased, there could be a critical, $N_*$, such that the corresponding $d_*$ becomes integer. It is, however, unclear whether an integer $N_*$ of this kind exists.

\subsection{Multi-trace scalars}

\begin{figure}[t!]
 \centering
 \begin{subfigure}[t]{0.3\textwidth}
      \centering
    \includegraphics[scale=0.3]{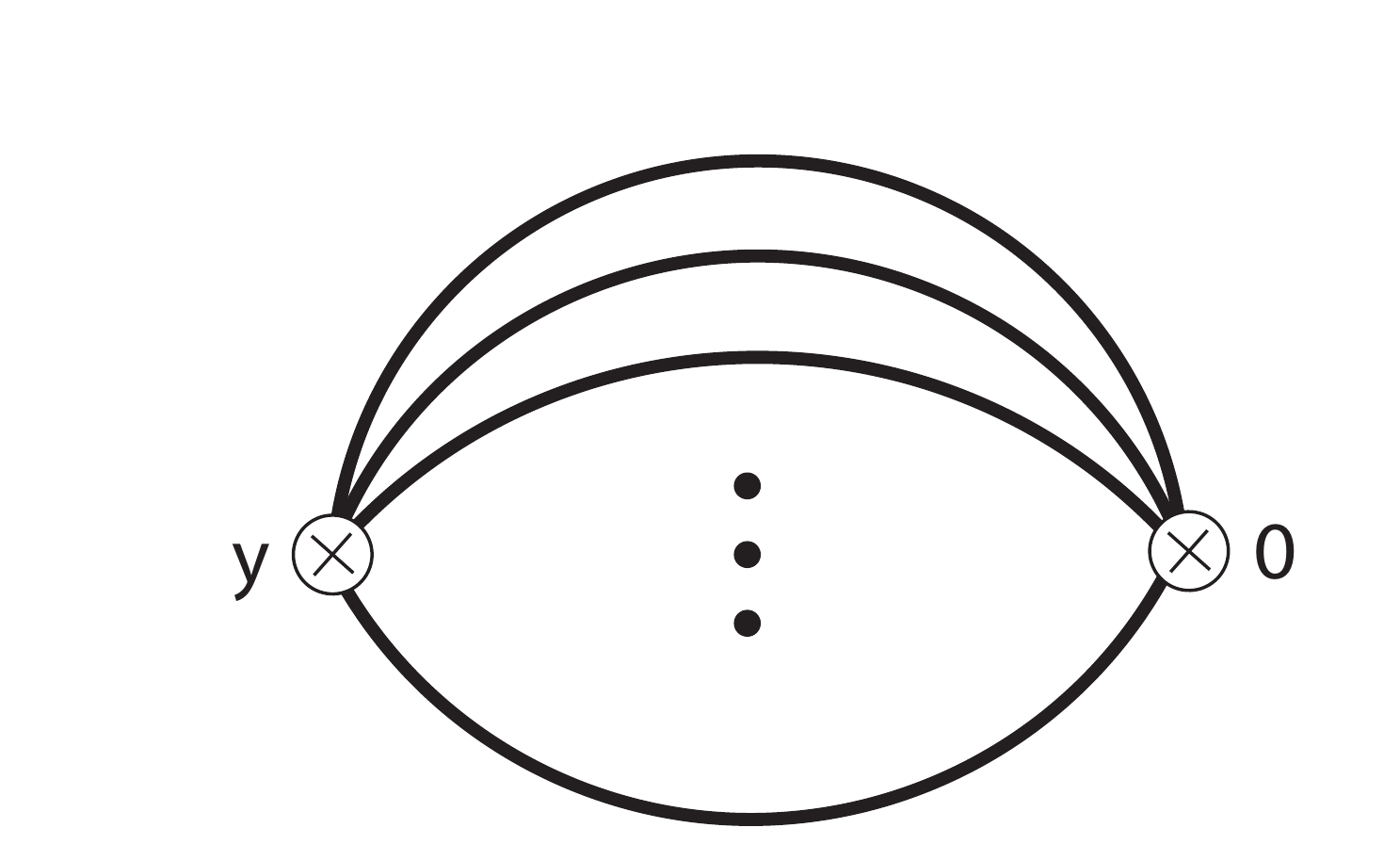}
  \caption{}    
    \end{subfigure}   
     \begin{subfigure}[t]{0.3\textwidth}
      \centering
    \includegraphics[scale=0.3]{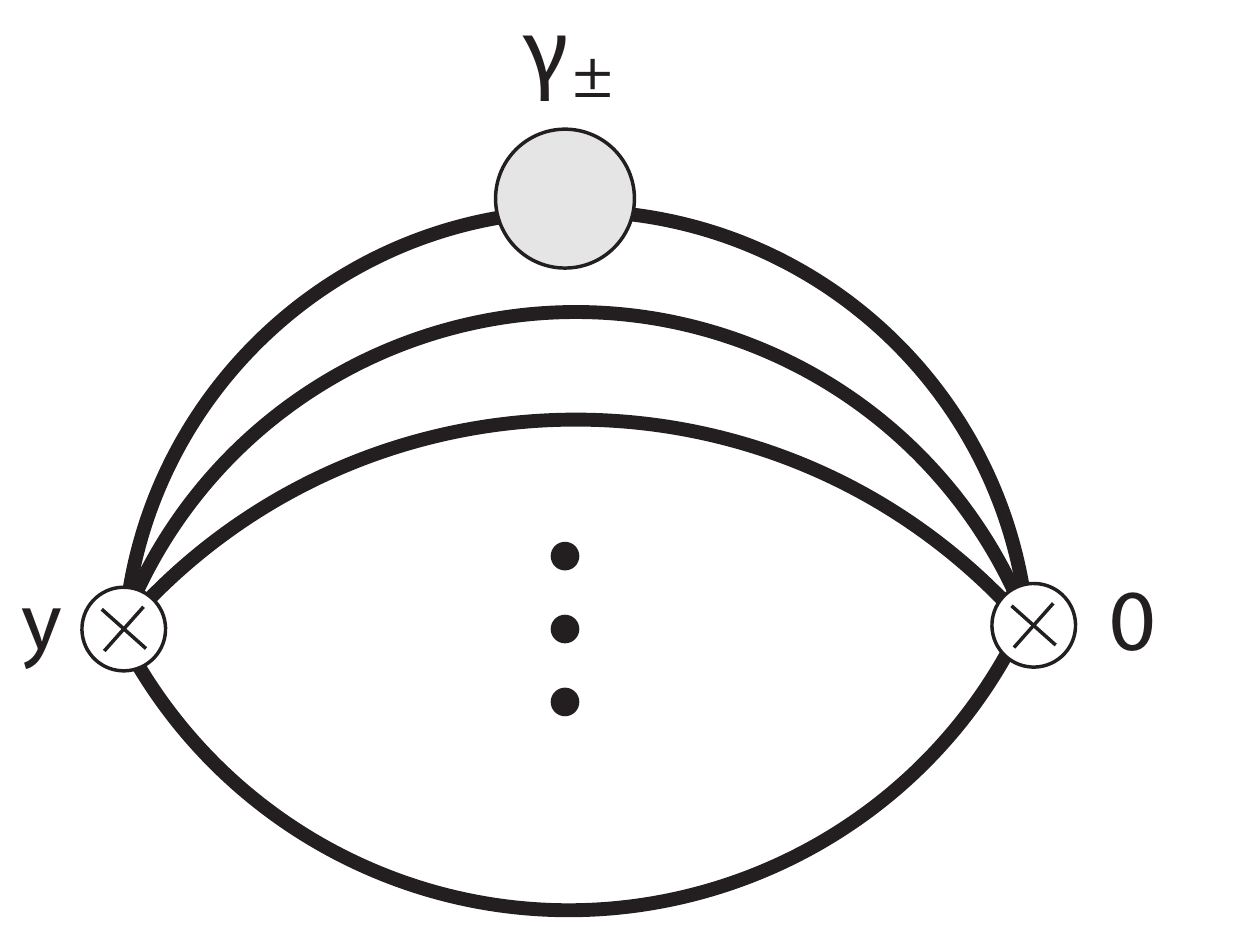}
  \caption{}    
    \end{subfigure}   
     \begin{subfigure}[t]{0.3\textwidth}
      \centering
    \includegraphics[scale=0.3]{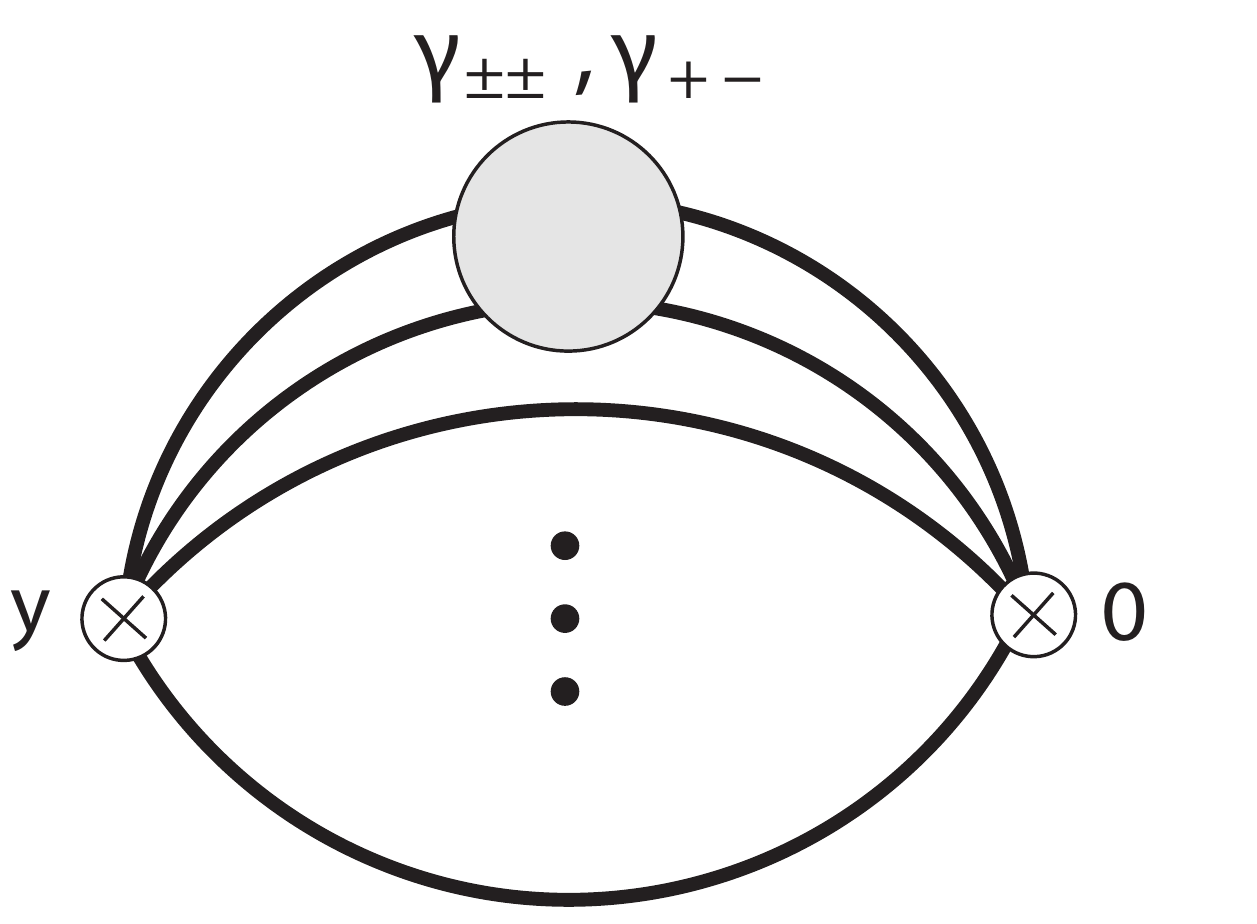}
  \caption{}    
    \end{subfigure}
    \caption{Feynman diagrams contributing to $\lan \mathcal{O}_{k,\ell} ~ \mathcal{O}_{k,\ell}\ran$ up to the next-to-leading order in $1/N$.  Cross-caps represent insertion points of $\mathcal{O}_{k,\ell}$. (a) Leading order correlator (b) $1/N$ correction associated with the single trace correlator (c) $1/N$ correction associated with the double trace correlator.}
    \label{mult-trace} 
\end{figure}

\noindent
The information about the single trace operators was crucial for the calculation of the anomalous dimensions of the double trace operators. However, this information was not sufficient. There were additional contributions, that needed to be accounted for while calculating the full answer. In contrast, the anomalous dimensions, $\gamma_{k,\ell}$, of the multi-trace scalar operators of the form $\mathcal{O}_{k,\ell}\equiv\sigma_+^k\sigma_-^\ell$, to leading order in $1/N$, are entirely determined by the diagrams that we evaluated already. That is to say, the complete answer can be written in terms of the anomalous dimensions of the single and double trace scalars. The latter can be readily understood based on the effective action analysis. 

\noindent
Indeed, $\gamma_{k,\ell}$ is defined by
\be
 \lan \mathcal{O}_{k,\ell}(y) ~ \mathcal{O}_{k,\ell}(0)\ran
 ={k! \, \ell! \, C_+^k \, C_-^\ell\big(1+A_{k,\ell}\big)\over |y|^{2(2k+\ell(d-2)+\gamma_{k,\ell})}}~,
\ee
where $A_{k,\ell}\sim 1/N$ represents a sub-leading correction to the amplitude. Now apart from the diagrams built off the effective cubic and quartic vertices in Fig.\ref{effective_vertices}, there are no other vertices which give rise to the $1/N$ corrections. In particular, as shown in Fig.\ref{mult-trace}, all possible diagrams, which contribute to the two point function of $ \mathcal{O}_{k,\ell}$ up to $1/N$ order, factorize into a product of zero order correlators $\lan\sigma_\pm \sigma_\pm\ran$ and previously calculated two point function of either single or double trace operators. Hence, calculating  $\gamma_{k,\ell}$ boils down to a simple problem in combinatorics, and the leading order anomalous dimension of $ \mathcal{O}_{k,\ell}$  takes the form
\be
 \gamma_{k, \ell}=(2-k-\ell)(k\gamma_+ + \ell\gamma_-) + k\ell \gamma_{+-} + {k(k-1)\over 2}\gamma_{++} 
 +{\ell(\ell-1)\over 2} \gamma_{--} ~.
 \label{anom-mult}
\ee

\noindent
Recall now that if the ranks of the groups are equal, \ie $m=N/2$, then the anomalous dimensions of the single and double trace scalars become simple functions of only the dimension $d$. Likewise, $\gamma_{k,\ell}$ simplify in this special case.

\section{Discussion}
\label{sec:discussion}

In this paper we studied the critical $O(m)\times O(N-m)$ vector model \eqref{action} in general $2\leq d\leq6$ dimensions. In the large $N$ limit and $\epsilon=4-d\ll 1$, this model exhibits a line of fixed points, which is lifted by the $1/N$ corrections provided that $\epsilon$ is the smallest expansion parameter in the problem \cite{Chai:2020onq,Chai:2020zgq}. To reveal what happens with  the conformal manifold in general dimension, we employed the large-$N$ technique to identify an operator which happens to be exactly marginal in any $d$ in the large-$N$ limit. Such an operator accompanies the existence of conformal manifold. We calculated the $1/N$ correction to its leading order scaling dimension and found that quite generically the anomalous dimension does not vanish, see \eqref{gamma_pm}, \eqref{gamma_pm_eqrank}. The outcome of this calculation is that only a discrete set of fixed points survive the $1/N$ corrections. An $O(N)$ invariant critical vector model is among them. This corresponds to the point on the conformal
manifold with $g_{11} = g_{22} = g_{12}$, in both the equal and unequal rank models. This model has been extensively studied throughout the literature for several decades, and we are not going to discuss it here. 

\noindent
We will discuss a new CFT with $g_{12}<0$. Note that in the equal rank model, this corresponds to the point $g_{11}=g_{22}=-g_{12}$ on the conformal manifold. In the unequal rank model, the critical couplings are a non-linear function of the ranks, and can only be determined numerically even in the perturbative regime near $d=4$. As shown in  \cite{Chai:2020onq,Chai:2020zgq}, it exhibits a persistent symmetry breaking at finite temperature. This is a rare example where a continuous global symmetry is broken at all temperatures for finite $N$. There is, however, a caveat here: this phenomenon occurs only in a fractional number of dimensions. While for infinite $N$ there is a candidate for a $3+1$ dimensional CFT with some of its internal symmetries broken at all temperatures \cite{Chaudhuri:2020xxb}, a fully fledged example of a finite $N$ CFT with persistent symmetry breaking in integer $d$ has not yet been found. For instance, the bi-conical model is free in $d=4$, whereas in $d=3$ it fails due to the Mermin-Wagner-Hohenberg-Coleman theorem \cite{Mermin:1966fe,Hohenberg:1967zz,Coleman:1973ci}.  Yet, understanding better the structure of such a CFT may help to either build a finite $N$ candidate with persistent symmetry breaking in realistic dimensions, {\eg} a model with persistent breaking of a discrete symmetry in $d=3$, or lead to a new no-go theorem forcing a mechanism for symmetry breaking restoration at finite $N$. 
To this end, we calculated a CFT data \eqref{anom-mult} associated with the scaling dimensions of the multi-trace scalar operators up to $1/N$ order. The dependence on the critical couplings is somewhat disguised 
in the closed form formulas, thereby allowing for explicit results only in $d=4-\eps$, where the coupling constants are perturbatively known. Away from the perturbative regime, the model becomes strongly coupled in the
IR, and the critical couplings cannot be evaluated anymore. However, there is a significant simplification in the limit of equal ranks, \ie for $m=N/2$. In this case we are able to obtain the scaling dimensions of the multi-trace scalar operators of the model as functions of only $d$ ($2\leq d\leq 6$).

\noindent
New analytical and numerical bootstrap techniques have been developed in recent times, many of which have been tested successfully in the $O(N)$ vector model at the Wilson-Fisher fixed point. The bi-conical vector model, for both equal and unequal ranks, can serve as a brand new testing ground for many of these bootstrap methods, which in turn could provide new conformal data for the model.

\noindent
While the model with equal ranks is a special case of \eqref{action}, there is no much difference between the equal and unequal ranks at zero temperature. Yet, the behaviour of the models is substantially different at finite temperature. In particular, if the ranks are equal and $N$ is finite, there is no persistent symmetry breaking in the equal rank case \cite{Chai:2020onq,Chai:2020zgq}. Hence, one might wonder whether a CFT data of the equal rank model may shed light on the structure of a theory with a persistent symmetry breaking. Hereafter we argue that the answer to this question is affirmative to a certain extent.

\subsection*{The slightly unequal rank model}

Let us introduce a new parameter $\delta=x-{1\over 2}$. It measures how far the model deviates from the equal rank case. The critical coupling constants of the conformal manifold, which survive the finite $N$ corrections, are functions of $\delta$, and we denote them by $h_{ij}(\delta)$. The equal rank model satisfies $h_{11}(0)=h_{22}(0)$, because the fixed points in this case respect the $\mathbb{Z}_2$ symmetry $\phi_1 \leftrightarrow\phi_2$, as can be explicitly checked using $\epsilon$-expansion \cite{Chai:2020onq,Chai:2020zgq}. Therefore we deduce that the following relations between the critical couplings hold
\be
h_{11}(\delta)=h_{22}(-\delta)~, \quad h_{12}^2(\delta)=h_{11}(\delta)h_{11}(-\delta) ~,
\ee
where the first identity rests on $x_1 \leftrightarrow x_2$ under $\delta\to -\delta$ (recall that $x_1={1\over 2} + \delta$ and $x_2={1\over 2} - \delta$), whereas the second one follows from $\det({\bf h})=0$. In particular, $h_{12}(\delta)$ is an even function of $\delta$, and all of the anomalous dimensions that we have calculated in this work depend on just one unknown function $h_{11}(\delta)$. Likewise, using these relations and \eqref{gamma_p}, \eqref{gamma_pp}, \eqref{gamma_mm}, \eqref{gamma_pm}, it can be readily verified that the leading order anomalous dimensions are also even functions of $\delta$.

\noindent
In particular, for a {\it slightly} unequal rank characterized by $\delta\ll 1$, various anomalous dimensions in general dimension $d$ satisfy\footnote{Assuming $g_{12}<0$ and $\epsilon=4-d\ll1$, one can verify it explicitly within $\epsilon$-expansion,
\begin{align}
 &\g_+ = {-2\eps(1 + 8\delta^2)\over N}+\ldots,\hspace{.3cm}\g_-={2\eps(3 -8\delta^2)\over N} + \ldots \nn\\
 &\g_{++}={-\eps^2(1 - 32\delta^2)\over N} + \ldots ,\hspace{.2cm}
 \g_{--}={12\eps-13\eps^2 + 16\eps(3+2\eps)\delta^2\over N} + \ldots\nn\\
 &\g_{+-}={16\eps (1 -6\delta^2)\over N} +\ldots
\end{align}
where ellipsis encode higher order terms in $1/N, \delta$ and $\epsilon$.
}
\be
 \gamma(\delta) = \gamma(0) + \mathcal{O}(\delta^2/N, 1/N^2 ) ~,
\ee 
where $\gamma(0)$ collectively represents the anomalous dimensions of the equal rank model, \eg it can be either \eqref{sing_anom}, \eqref{eqrank gamma_pp},  \eqref{eqrank gamma_mm} or \eqref{gamma_pm_eqrank}. 

\noindent
In contrast, the order parameter of the symmetry breaking scales linearly with $\delta\ll 1$ at any temperature \cite{Chai:2020onq,Chai:2020zgq}, implying that to linear order in $\delta$ a CFT data of the unequal rank model matches that of an equal rank case, albeit for unequal ranks the system already exhibits a pattern of persistent symmetry breaking. Moreover, in these models, the non-zero value of the marginal operator to the leading order in $\delta$ indicates
that the conformal manifold is lifted. We cannot, however, determine the fate of the conformal manifold
in the most general unequal rank model due to lack of knowledge of the critical couplings. \\
As an example, consider $N = 10$ and $\delta = 0.1$. This corresponds to an $O(4) \times O(6)$ model, which can
be classified as a slightly unequal rank model. In this model, the anomalous dimension of the marginal
operator will be given for any $2 < d < 6$ by \eqref{gamma_pm_eqrank} $+O(\delta^2)$ corrections, indicating that the conformal
manifold is lifted. The same cannot be said, however, for example in an $O(1) \times O(9)$ model with $\delta = 0.4$, where the leading anomalous dimension of the marginal operator itself is dependent on the critical couplings.

\subsection*{A non-trivial fixed point in $d=3$} 

Our calculations provide an additional evidence for the existence of an interesting strongly interacting CFT in three dimensions. Most illuminating in this regard is the equal rank model, where we evaluated the numerical values of the scaling dimensions of various operators. 

\noindent
Tractable CFTs in three dimensions are interesting from a number of perspectives. The critical $O(N)$ vector model in $d=3$ presents the most prominent example in this regard. It has a perturbative IR fixed point in $d=4-\eps$ dimensions, which is the well known {\it Wilson-Fisher} fixed point. In fact, the $\epsilon$-expansion was devised to gain a quantitative description of a fixed point in $d=3$, which describes the three-dimensional critical Ising model. The main difficulty encountered during the studies of the Ising model stems directly from its strongly coupled nature, and not until very recently (see \cite{Simmons-Duffin:2016wlq} and \cite{Caron-Huot:2020ouj} for recent progress) much was unknown about it due to insufficient non-perturbative techniques. 

\noindent
Nonetheless, one can study the model in the limit of large $N$ and make much progress in characterising the low-lying spectrum of the theory. The primary reason for the effectiveness of the large-$N$ approach is due to the coupling constant dropping out of the analysis, thereby ridding us off the necessity to determine its critical value, which is beyond the reach of analytical approach. 

\noindent
This entire discussion carries over to the equal rank model (and in some limited way to a slightly unequal rank case). Thus, for instance, a perturbative analysis of the bi-conical model in the vicinity of $d=4$ results in the Wilson-Fisher like fixed points. Moreover, due to the $\mathbb{Z}_2$ symmetry $\phi_1 \leftrightarrow\phi_2$, the large $N$ analysis of the equal rank bi-conical model reveals that we can extend the calculations beyond the perturbative regime in the vicinity of $d=4$.\footnote{This is not true for the unequal rank model where there is no symmetry to simplify the calculations.} 

\noindent
The punchline of this analysis is that very much like the case of the critical $O(N)$ vector model, the values of the couplings are not needed to derive the large-$N$ CFT data for the equal rank bi-conical model. 

\noindent
Moreover, to leading order in the large-$N$ expansion the bi-conical model disintegrates into two free scalar fields $\phi_1$ and $\phi_2$ in the vector representation of $O(m)$ and $O(N-m)$ respectively, and decoupled scalar singlets of $O(m)\times O(N-m)$ with scaling dimension $\Delta_+=2$ and $\Delta_-=d-2$. The sub-leading $1/N$ terms re-instate the interaction between various fields, thereby giving us a non-trivial interacting model in $d=3$.

\section*{Acknowledgements} \noindent RS dedicates this paper to the loving memory of his father Dipak Kumar Sinha. We thank Soumyadeep Chaudhuri, Changha Choi, Mikhail Goykhman and Zohar Komargodski for helpful discussions. Our work is partially supported by the Israeli Science Foundation Center of Excellence (grant No. 2289/18). The work of NC, RS and MS is partially supported by the Binational Science Foundation (grant No. 2016186),  and by the Quantum Universe I-CORE program of the Israel Planning and Budgeting Committee (grant No. 1937/12). 

\appendix

\section{Beta functions}
\label{appx: beta function}

In this Appendix we consider Wilsonian RG flow of the model \eqref{action} to derive the beta functions of the couplings. The partition function at some arbitrary scale $\mu$ is given by
\be
Z_\mu=\int \prod_{a=1}^2 D\phi_a  \exp\[ - \sum_{i=1}^2 \int d^dx ~ \( \frac{1}{2}\partial_\nu \phi_i \partial^\nu \phi_i +\frac{1}{2} g_{ i}^2 \, \mu^2 \phi^2_i+\frac{g_{ij}}{N} \mu^{4-d} \,  \phi^2_i\,  \phi^2_j + \ldots\) \] \, ,
\ee
where ellipsis encode $\phi^6$ and higher order interactions with or without derivatives. Note that in contrast to the main body of the text, the coupling constants $g_{ij}$ are dimensionless throughout this Appendix. 

\noindent
The interaction matrix $g_{ij}$ is symmetric, and every principal submatrix of $g_{ij}$ is positive definite to make sure the theory is stable for large values of the fields. However, this is not essential for the calculations carried out in this Appendix.

\noindent
To find the fixed points, one has to analyze an infinite system of one-loop exact beta functions. Unfortunately, it cannot be done in full generality. However, assuming $d=4-\epsilon$ with $|\epsilon|\ll 1$, we can implement the classic Wilson-Fisher $\epsilon$-expansion. In this case apart from $\phi^4$ and $(\del\phi)^2$, the coupling parameter of any interaction with $\ell$ factors of the fields will have a power series expansion which begins with a term of order $\epsilon^{\ell/2}$. Hence, to linear order in $\epsilon$ we can focus on $g_{i}$ and $g_{ij}$ only.

\noindent
Integrating out field modes with energy between $(\mu, \mu-d\mu)$, yields\footnote{No summation over repeated indices in what follows.}
\bea
&&g_{i}(\mu-d\mu) (\mu-d\mu)^2=g_{i}(\mu) \mu^2 + 4 \sum_k g_{ik} \mu^{4-d}
 \(x_k +{2\over N} \delta_{ik}\)
\int_{\mu-d\mu}^\mu {d^dp\over (2\pi)^d} {1\over p^2 + \mu^2 g_{k}(\mu)} ~,
\non\non
&& g_{ij}(\mu-d\mu) (\mu-d\mu)^{4-d}= g_{ij}(\mu) \mu^{4-d}
\label{RGflow}
\\
&&
-4\sum_k g_{ik}g_{jk} \mu^{2(4-d)} \(x_k+{2\over N}(\delta_{ik}+\delta_{jk})\)
\int_{\mu-d\mu}^\mu {d^dp\over (2\pi)^d} {1\over \Big( p^2 + \mu^2 g_{k}(\mu) \Big)^2}
\non
&&
-{16\over N}g_{ij}^2  \mu^{2(4-d)} 
\int_{\mu-d\mu}^\mu {d^dp\over (2\pi)^d} 
{1\over \Big( p^2 + \mu^2 g_{i}(\mu) \Big)\Big( p^2 + \mu^2 g_{j}(\mu) \Big)}~.
\nonumber
\eea
The loop integrals are simple since they are done over a thin shell,
\bea
 \mu {dg_{i}\over d\mu} &=& - 2 g_{i} - {S_d\over (2\pi)^d} \sum_k 
 \Big(x_k+{2\over N} \delta_{ik}\Big) {4 g_{ik} \over 1 + g_{k}} ~,
 \\
 \mu {dg_{ij}\over d\mu} &=& -\epsilon \, g_{ij} + {4S_d\over (2\pi)^d} 
 \sum_k \Big(x_k+{2\over N}(\delta_{ik}+\delta_{jk})\Big) {g_{ik} g_{jk}\over (1+g_{k})^2}
 \nonumber \\
 &+&{16\over N}  {S_d\over (2\pi)^d}  {g_{ij}^2  \over (1+g_{i})^2(1+g_{j})^2} +\mathcal{O}(\epsilon^3)~,
 \nonumber
\eea
where $S_d={2\pi^{d/2}\over \Gamma(d/2)}$. Since $g_{i}\sim \epsilon$, one can replace all the denominators in the above equation by 1 without violating the accuracy. Hence, we finally get
\bea
 \mu {dg_{i}\over d\mu} &=& - 2 g_{i} - a \sum_k 
 x_k g_{ik} -{2 a\over  N} g_{ii} + \mathcal{O}(\epsilon^2)~,
 \label{Wil_beta_func}\\
 \mu {dg_{4ij}\over d\mu} &=& -\epsilon \, g_{ij} + 
 a \sum_k \Big(x_k+{2\over N}(\delta_{ik}+\delta_{jk})\Big) g_{ik} g_{jk}
 + { 4 a \over  N} g_{ij}^2   + \mathcal{O}(\epsilon^3)~.
 \nonumber
\eea
where we introduced $a=4 S_d/(2\pi)^d=1/(2\pi^2)+\mathcal{O}(\epsilon)$. The fixed points $g_{i}^*, g_{ij}^*$ of the flow satisfy
\bea
 g_{i}^* &=& - {a\over 2}  ( x_1 \, g_{i1}^* + x_2 \, g_{i2}^*) - {a\over   N} g_{ii}^*  + \mathcal{O}(\epsilon^2)~,
 \\
 0 &=& -\epsilon \, g_{ij}^* + a \(  x_1 \, g_{i1}^* g_{j1}^* + x_2 \, g_{i2}^* g_{j2}^*  \) 
 + {2 a\over N} \(  g_{ii}^* +g_{jj}^*\)g_{ij}^* 
  + { 4 a \over  N} g_{ij}^{*2}+ \mathcal{O}(\epsilon^3)~.
 \nonumber
\eea
The above algebraic equations are reliable up to linear order in $\epsilon$. After solving for $g_{ij}^*$ one substitutes it into the first equation to get $g_i^*$. We ignore equation for $g_i^*$ since it is not essential for our needs. For $g_{12}^*\neq 0$ the rest reduces to
\bea
 \epsilon g_{11}^*&=&a\(x_1+{8\over N}\)g_{11}^{*\,2} + a \, x_2 \,g_{12}^{*\,2}~,
 \non
  \epsilon g_{22}^*&=&a\(x_2+{8\over N}\)g_{22}^{*\,2} + a \, x_1 \,g_{12}^{*\,2}~,
  \non
 \epsilon &=& a\(x_1  + {2 \over N} \) g_{11}^* + a \(x_2 + {2 \over N} \)g_{22}^*+ { 4 a \over  N} g_{12}^{*}~.
\eea
These equations are degenerate in the limit $N\to\infty$, and there is a line of fixed points
\be
 \epsilon = a\big(x_1 g_{11}^* + x_2 g_{22}^*\big) ~, \quad \text{det}(g_{ij}^*)=0~.
\ee

\section{Conformal perturbation theory}
\label{appx:identities}

There are simple diagrammatical rules for performing some of the loop integrals within conformal perturbation theory. In this Appendix we review and apply these rules to evaluate Feynman diagrams which contribute to the anomalous dimensions studied in this paper. The calculations are done in position space. For simplicity, we assume that in any Feynman graph the interaction vertices and amplitudes of the propagators are normalized to unity, \eg

\begin{center}
  \begin{picture}(156,23) (31,-11)
  \thicklines
   \put(40,-2){\line(1,0){88}}
    \put(40,-2){\circle*{4}}
    \put(128,-2){\circle*{4}}
    \Text(78,0)[lb]{\scalebox{0.8}{$2\Delta$}}
    \Text(152,-15)[lb]{\scalebox{1.2}{$=\frac{1}{|x|^{2\Delta}}$}}
  \end{picture}
\end{center}

\noindent
In particular, a simple loop diagram satisfies additivity, \ie

\begin{figure}[H]
\centering \noindent
\hfill\includegraphics[width=14cm]{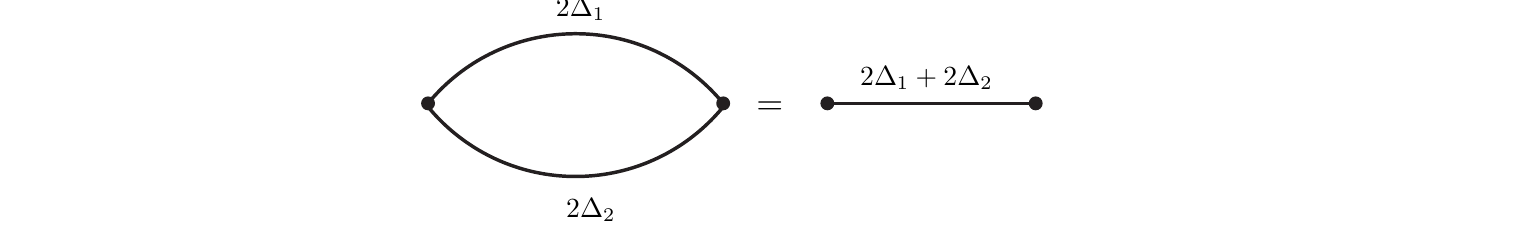}\hspace*{\fill}

\label{fig:loop additivity}
\end{figure}

\noindent
Furthermore, there is a propagator merging relation of the form
\begin{equation}
\label{propagator merging relation}
 \int d^d x_2 ~ {1\over |x_{12}|^{2\Delta_1} |x_{23}|^{2\Delta_2}}= {U(\Delta_1,\Delta_2,d-\Delta_1-\Delta_2)\over |x_{13}|^{2(\Delta_1+\Delta_2)-d}} ~,
\end{equation}
where 
\begin{equation}
U(\Delta_1,\Delta_2,\Delta_3) = \pi^\frac{d}{2}\,\frac{\Gamma\left(\frac{d}{2}-\Delta_1\right)\Gamma\left(\frac{d}{2}-\Delta_2\right)\Gamma\left(\frac{d}{2}-\Delta_3\right)}
{\Gamma(\Delta_1)\Gamma(\Delta_2)\Gamma(\Delta_3)}\,.
\end{equation}
This relation can be represented diagrammatically as follows

\begin{figure}[H]
\centering \noindent
\hfill\includegraphics[width=14cm]{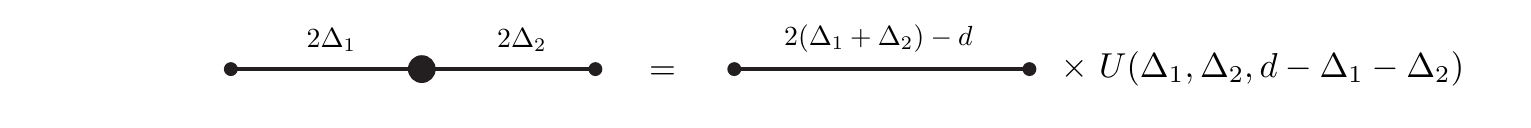}\hspace*{\fill}

\label{fig:propagator merging}
\end{figure}

\noindent
where we integrate over the insertion point of the bold vertex on the left-hand side.

\noindent
Finally, the star-triangle relation for the conformal cubic vertex is given by,
\begin{equation}
\label{uniqueness scalar}
\int d^dx_4\,\frac{1}{|x_{14}|^{2\Delta_1}|x_{24}|^{2\Delta_2}|x_{34}|^{2\Delta_3}}
= \frac{U (\Delta_1,\Delta_2,\Delta_3)}{|x_{12}|^{d-2\Delta_3}
|x_{13}|^{d-2\Delta_2}|x_{23}|^{d-2\Delta_1}} ~ , \quad \Delta_1+\Delta_2+\Delta_3 = d ~.
\end{equation}
Diagrammatically it can be represented as follows

\begin{figure}[H]
\centering \noindent
\hfill\includegraphics[width=12cm]{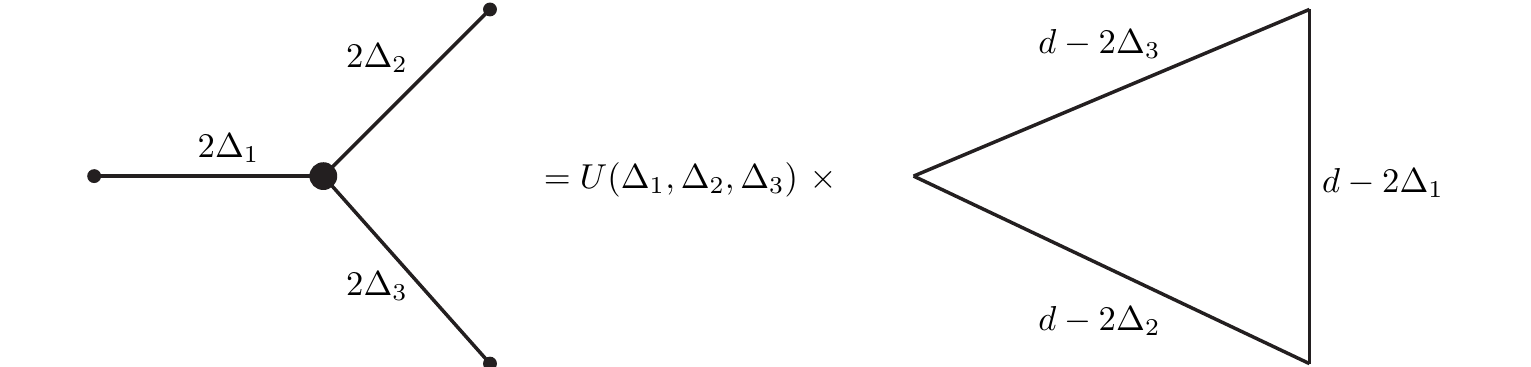}\hspace*{\fill}

\label{fig:star triangle identity.}
\end{figure}

\noindent
Let us illustrate the power of these rules by evaluating the diagrams contributing to the anomalous dimensions of various operators studied in our work. 

\begin{figure}[H]

    \begin{subfigure}[t]{1.05\textwidth}

    	\includegraphics[scale=0.45]{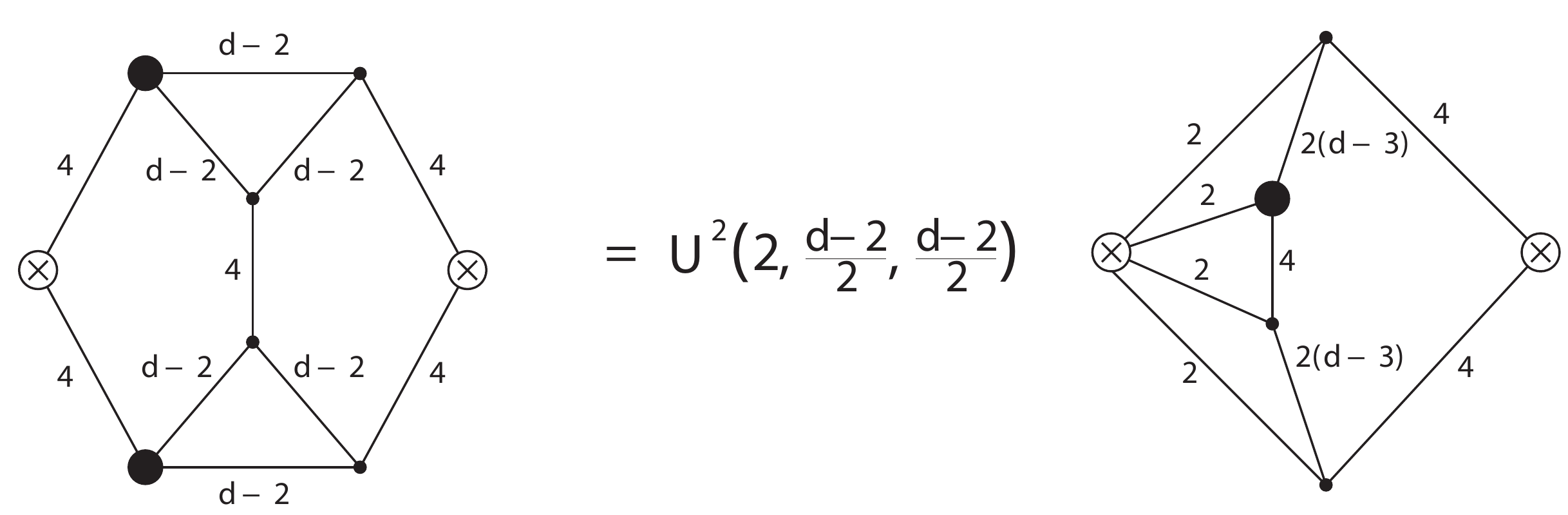} 

    \end{subfigure}
   	 \hfill
    \begin{subfigure}[t]{1.05\textwidth}
    	\includegraphics[scale=0.45]{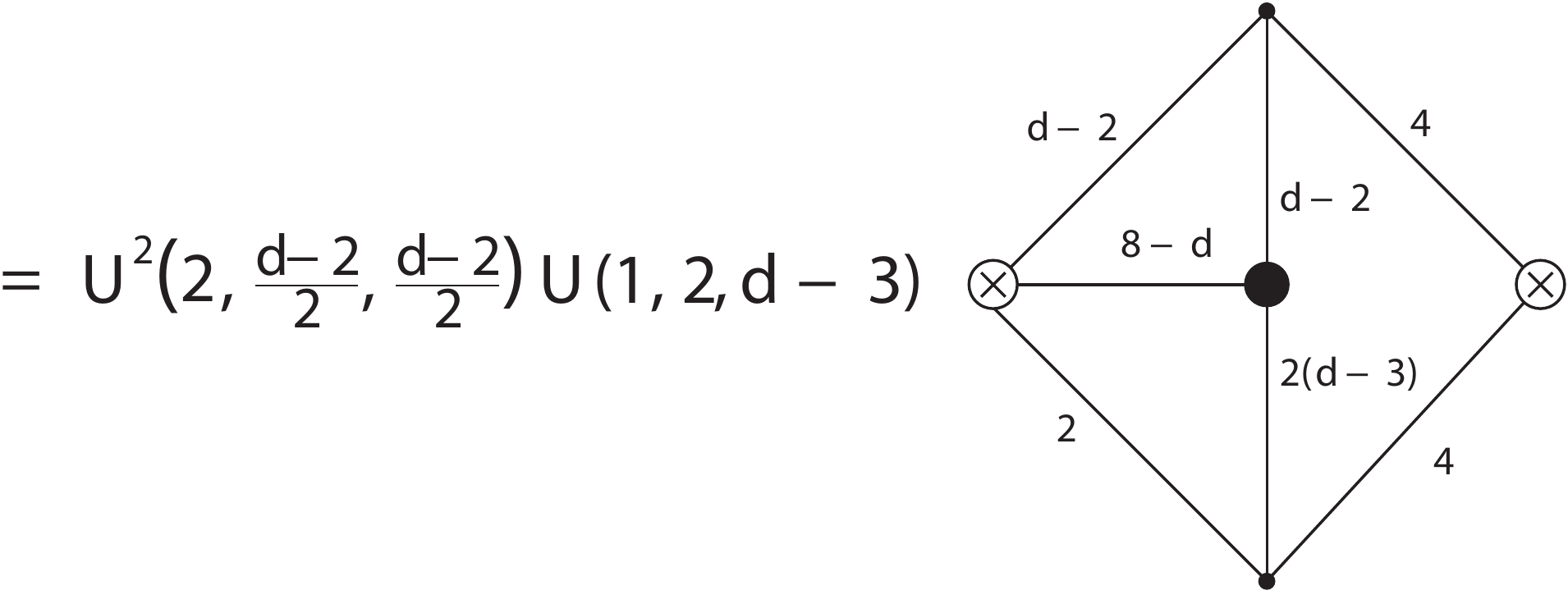}
    \end{subfigure}
    	\hfill
    \begin{subfigure}[t]{1.05\textwidth}
     	\includegraphics[scale=0.45]{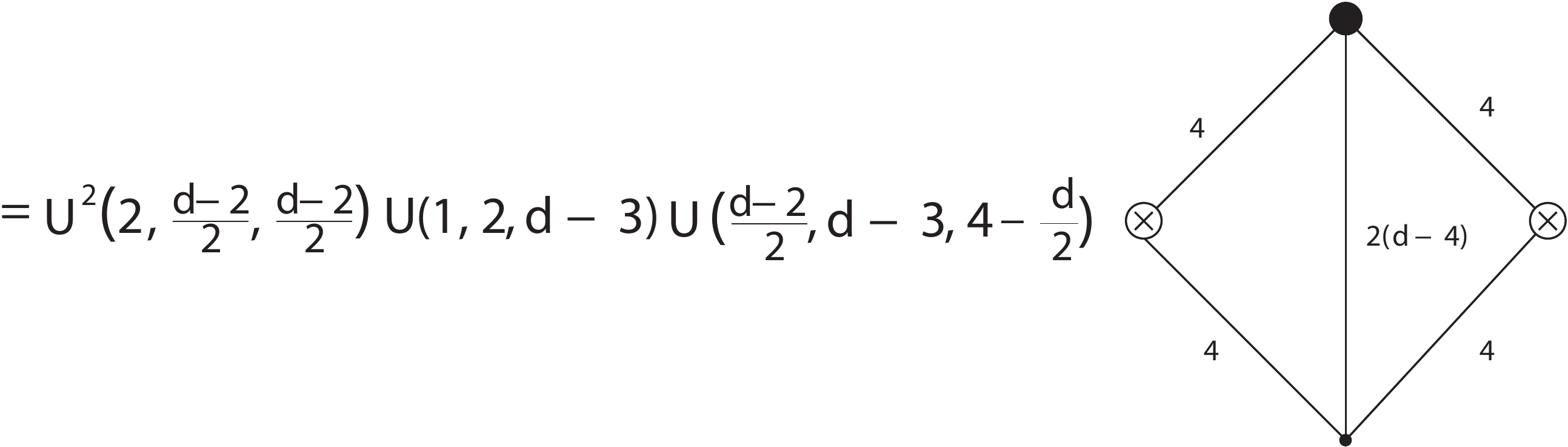}
    \end{subfigure}
    	\hfill
    \begin{subfigure}[t]{1.05\textwidth}
     	\includegraphics[scale=0.45]{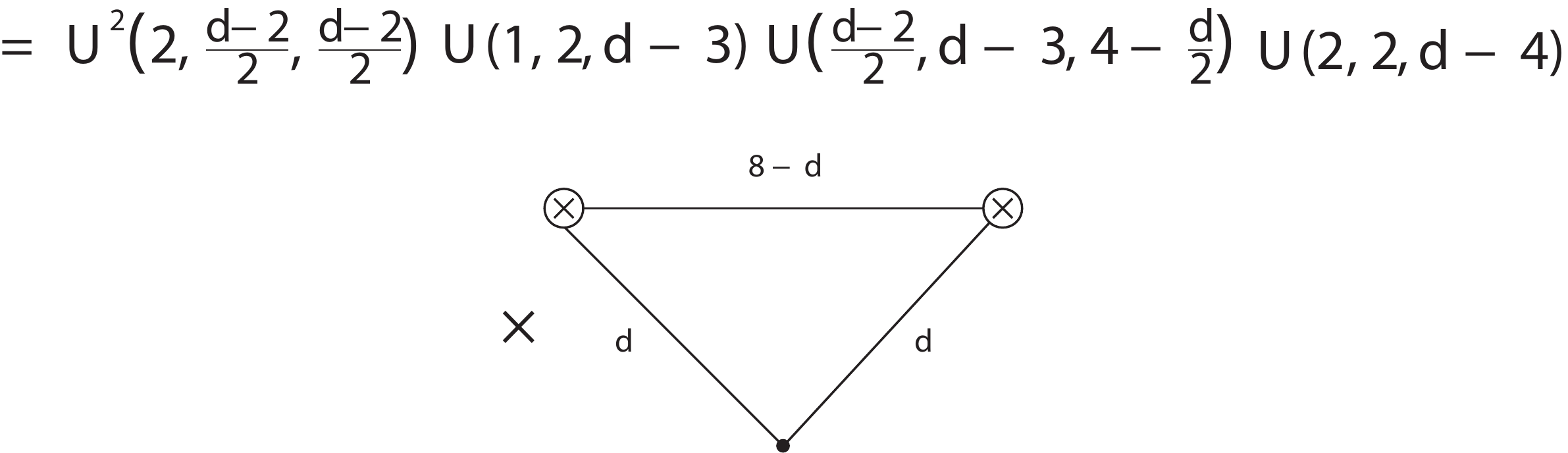}
    \end{subfigure}
        \caption{Calculation of the loop integrals associated with the diagram in Fig.\ref{sigma_dt-1}(c). In each step the star-triangle relation \eqref{uniqueness scalar} is applied to integrate over the insertion point of the bold vertex. Cross-caps denote location of the external operators.}
    \label{sigpp-c}
 \end{figure}

\noindent
As shown in Fig.\ref{sigpp-c}, the star-triangle relation \eqref{uniqueness scalar} is enough to evaluate the logarithmically divergent part of the diagram in Fig.\ref{sigma_dt-1}(c). 
\noindent
The remaining integral in Fig.\ref{sigpp-c} diverges. Introducing a spherical sharp cut off, $\mu$, yields
 \be
  \text{Fig}.\ref{sigpp-c}={16 \pi^{3d} (d-2) \Gamma^2\({2-d\over 2}\)\over (d-4)^3 \, \Gamma^2(d-3)} {\log\,\mu\over |y|^8} ~.
  \label{figpp_c}
 \ee

\begin{figure}[H]

    \begin{subfigure}[t]{1\textwidth}

    	\includegraphics[scale=0.5]{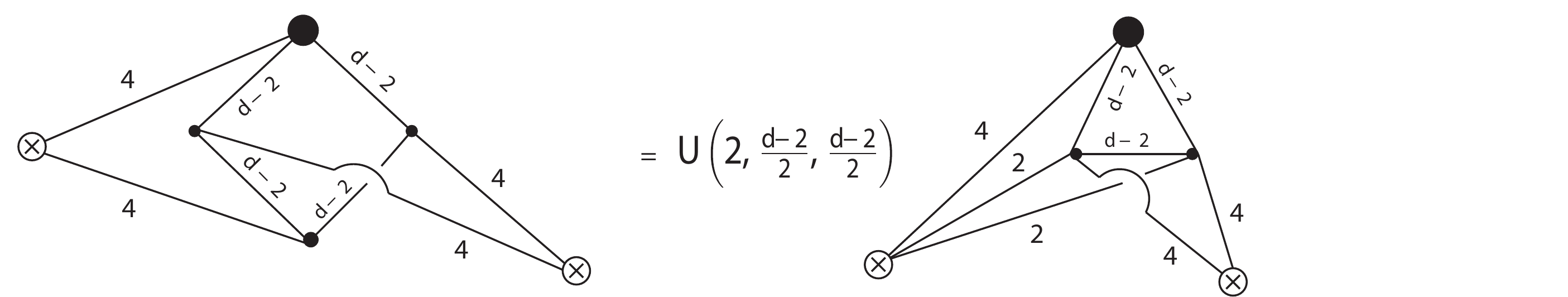} 

    \end{subfigure}
   	 \hfill
    \begin{subfigure}[t]{1\textwidth}
    	\includegraphics[scale=0.5]{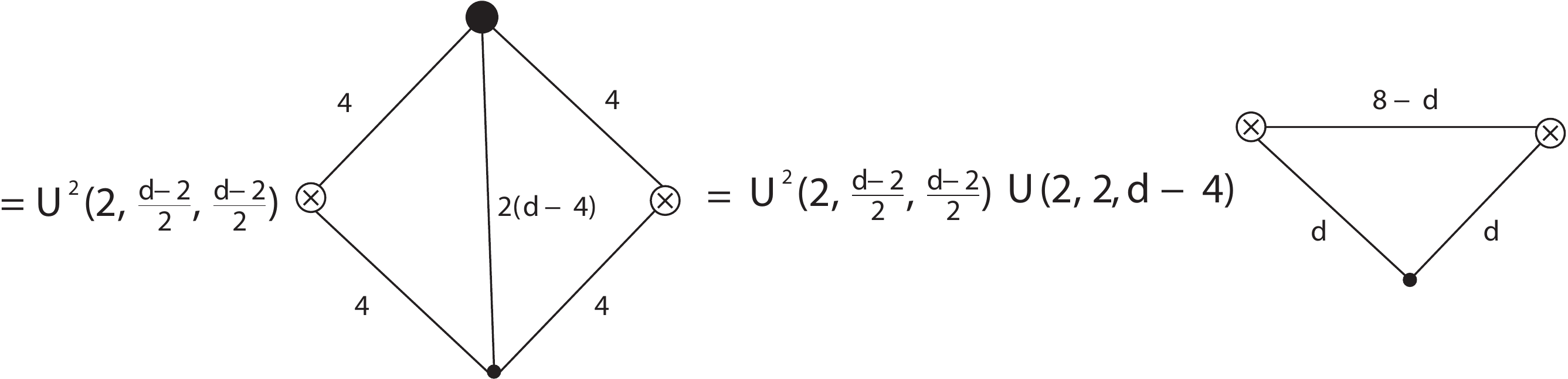}
    \end{subfigure}
        \caption{Calculation of the integrals associated with the diagram in Fig.\ref{sigma_dt-1}(e). In each step the star-triangle relation \eqref{uniqueness scalar} is applied to integrate over the insertion point of the bold vertex. Cross-caps denote location of the external operators.}
    \label{sigpp-e}
 \end{figure}

\noindent
Similarly, one can evaluate the diagrams in Fig.\ref{sigma_dt-1}(d,e). For instance, the necessary steps for the diagram in Fig.\ref{sigma_dt-1}(e) are shown explicitly in Fig.\ref{sigpp-e}.
\noindent
As before, introducing a spherical cut off $\mu$ to carry out the remaining integral in \text{Fig}.\ref{sigpp-e}, yields
 \be
  \text{Fig}.\ref{sigpp-e}={ (8\pi^d)^2 \Gamma\({8-d\over 2}\)\over (d-4)^3 \, \Gamma(d-3)\Gamma\({d\over 2}\)} {\log\,\mu\over |y|^8} ~.
  \label{figpp_e}
 \ee

\noindent
Recall now that the non-trivial contribution to $\gamma_{--}$ comes entirely from Fig.\ref{sigma_dt-2}(b,c). Fig.\ref{sigmm-c} below summarizes all the steps which are necessary to calculate the diagram in Fig.\ref{sigma_dt-2}(c).
\begin{figure}[H]
\centering
    	\includegraphics[scale=0.5]{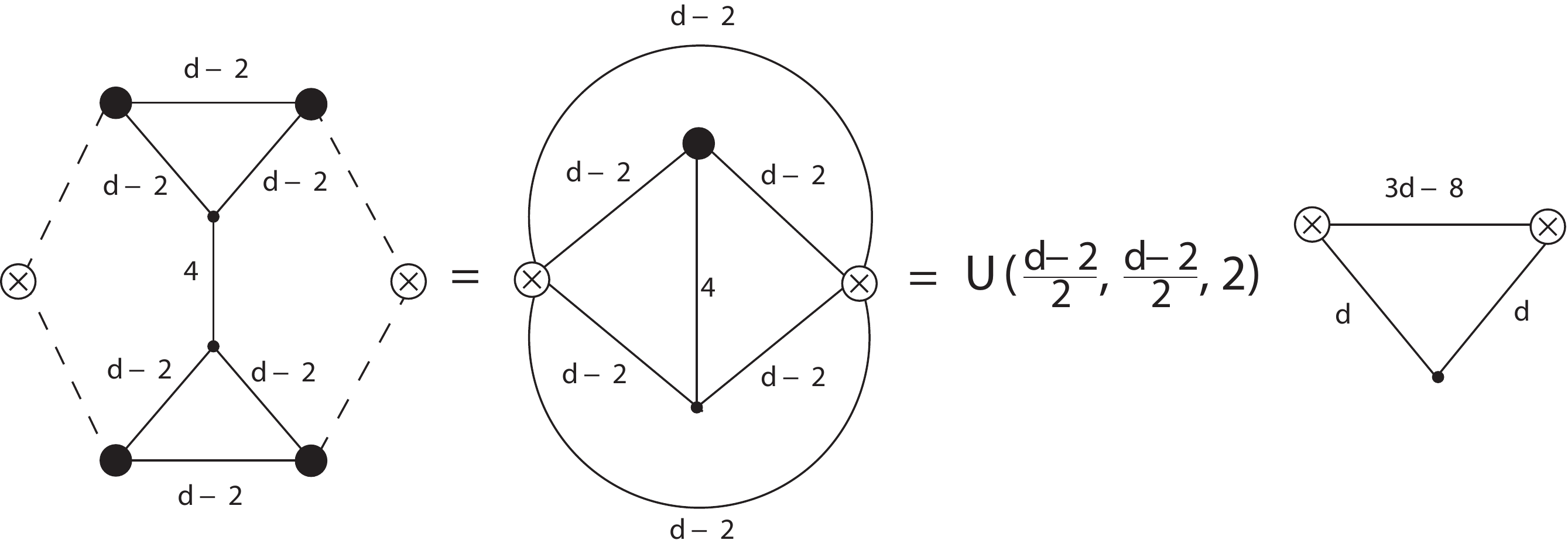} 
        \caption{Calculation of the loop integrals associated with the diagram in Fig.\ref{sigma_dt-2}(c). The dashed lines represent a delta function of the $\sigma_-$ propagator, and we integrate over the delta functions first. Next the star-triangle relation \eqref{uniqueness scalar} is applied to integrate over the insertion point of the bold vertex. Cross-caps denote location of the external operators.}
    \label{sigmm-c}
 \end{figure}
\noindent
In terms of spherical cut off $\mu$, we get
\be
  \text{Fig}.\ref{sigmm-c}={ 4(d-2)\pi^d \over (d-4) \, \Gamma^2\({d\over 2}\)} {\log\,\mu\over |y|^{4(d-2)}} ~.
  \label{figmm_c}
\ee

\begin{figure}[H]

    \begin{subfigure}[t]{1\textwidth}

    	\includegraphics[scale=0.5]{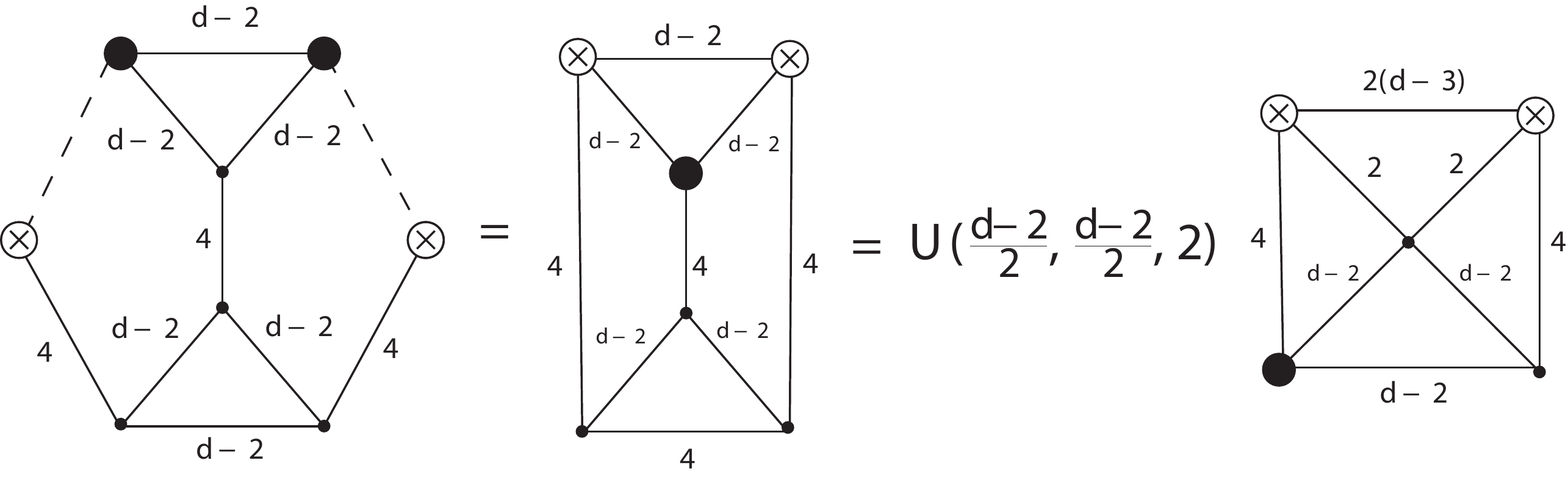} 

    \end{subfigure}
   	 \hfill
    \begin{subfigure}[t]{1\textwidth}
    	\includegraphics[scale=0.5]{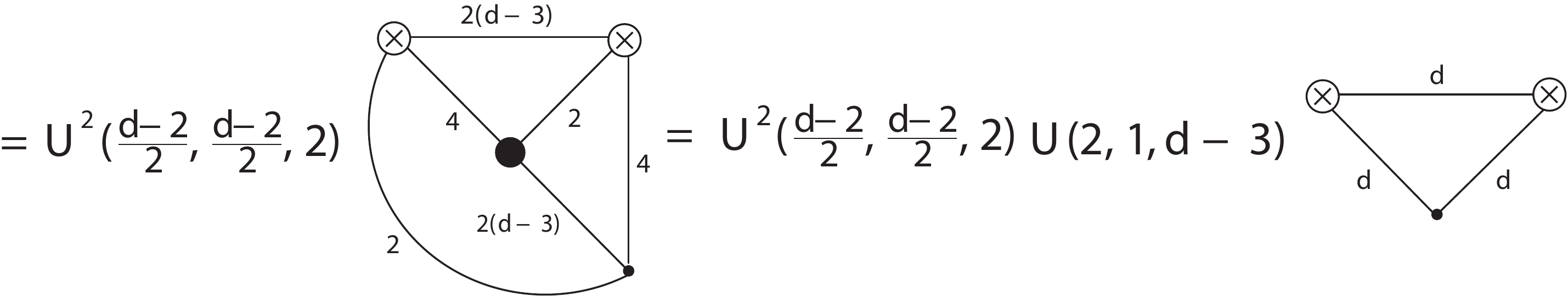}
    \end{subfigure}
        \caption{Calculation of the integrals associated with the diagram in Fig.\ref{sigma_dt-3}(c). The dashed lines represent a delta function of the $\sigma_-$ propagator, and we integrate over the delta functions first. The star-triangle relation \eqref{uniqueness scalar} is applied to integrate over the insertion point of the bold vertex for each step thereafter. Cross-caps denote location of the external operators.}
    \label{sigpm-c}
 \end{figure}

\noindent
Next let us consider $\gamma_{+-}$. All relevant diagrams for the calculation of this anomalous dimension are listed in Fig.\ref{sigma_dt-3}. It turns out that the same calculation should be done to evaluate the loop integrals of the graphs shown in Fig.\ref{sigma_dt-3}(c,d). It is displayed in Fig.\ref{sigpm-c}. 
 \noindent
 Equivalently,
 \be
  \text{Fig}.\ref{sigpm-c}={ - 16 \pi^{2d} \Gamma\({4-d\over 2}\)\over (d-4)^2 \, \Gamma(d-3)\Gamma\({d\over 2}\)} {\log\,\mu\over |y|^{2d}} ~.
  \label{figpm_c}
 \ee

\noindent
Finally, to evaluate Fig.\ref{sigma_dt-3}(g) one should follow the steps in Fig.\ref{sigpm_g} below.
\begin{figure}[H]
\centering
    	\includegraphics[scale=0.5]{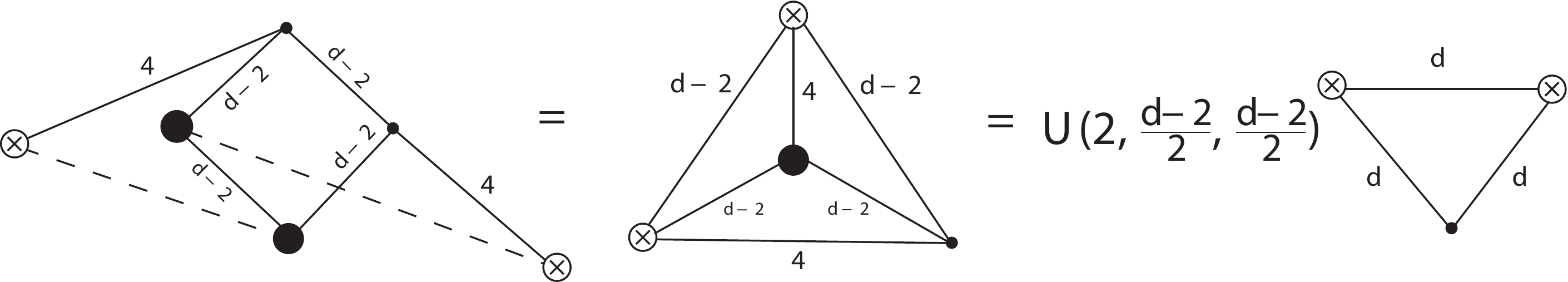} 
        \caption{Calculation of the loop integrals associated with the diagram in Fig.\ref{sigma_dt-3}(g). The dashed lines represent a delta function of the $\sigma_-$ propagator, and we integrate over the delta functions first. The star-triangle relation \eqref{uniqueness scalar} is applied to integrate over the insertion point of the bold vertex in the last equality. Cross-caps denote location of the external operators. }
    \label{sigpm_g}
 \end{figure}
 \noindent
 Up to a minor change in the scaling of the figure, the final answer is identical to \eqref{figmm_c}
\be
  \text{Fig}.\ref{sigpm_g}={ 4(d-2)\pi^d \over (d-4) \, \Gamma^2\({d\over 2}\)} {\log\,\mu\over |y|^{2d}} ~.
  \label{figpm_g}
\ee

\bibliographystyle{utphys}
\bibliography{ref} 

\end{document}